\documentclass[apj]{emulateapj}

\usepackage{graphicx}
\usepackage{rotating}
\usepackage{afterpage}
\usepackage{txfonts}
\usepackage[]{natbib}
\bibpunct{(}{)}{;}{a}{}{,}

\newcommand{\um}{$\mu$m}

\newcommand{\kms}{km\thinspace s$^{-1}$}

\def\arcmin{\hbox{$^\prime$}}
\def\arcsec{\hbox{$^{\prime\prime}$}}
\def\utw{\smash{\rlap{\lower5pt\hbox{$\sim$}}}}
\def\udtw{\smash{\rlap{\lower6pt\hbox{$\approx$}}}}

\def\Lsun{\hbox{\it L$_\odot$}}

\def\Teff{\hbox{\it T$_{\rm eff}$}}

\def\Msun{\hbox{\it M$_\odot$}}

\def\Mbol{\hbox{\it M$_{bol}$}}

\def\Teff{\hbox{\it T$_{\rm eff}$}}

\def\Mk{\hbox{\it M$_{\rm K}$}}
\newcommand{\Ks}{{\it K$_{\rm s}$}}

\newcommand{\Aks}{{\it A$_{\rm K_{\rm s}}$}}

\newcommand{\Av}{{\it A$_{\rm V}$}}

\def\BCK{\hbox{\it BC$_{\rm K}$}}

\def\simgr{\mathrel{\hbox{\rlap{\hbox{\lower4pt\hbox{$\sim$}}}\hbox{$>$}}}}
\def\HH{H{\sc ii}}	

\def\brg{\hbox{ Br$_\gamma$}}

\def\Vlsr{\hbox{\it V$_{\rm LSR}$}}
\def\nar{New A Rev.}%

\shorttitle{Newly detected massive stars in the W33 complex.}
\shortauthors{Messineo et al.}

\begin{document}


\title{ Massive stars in the W33 giant molecular complex.}


\author{Maria~Messineo\altaffilmark{1,11}, 
        J. Simon Clark\altaffilmark{2},
        Donald~F.~Figer\altaffilmark{3},
        Rolf-Peter ~Kudritzki\altaffilmark{4,10},
	    Francisco Najarro\altaffilmark{5},
        R. Michael Rich\altaffilmark{6},
        Karl M. Menten\altaffilmark{1},
        Valentin D. Ivanov\altaffilmark{7,8},
	    Elena Valenti \altaffilmark{8},
	    Christine Trombley \altaffilmark{3},     
        C.-H. Rosie Chen    \altaffilmark{1}, 
	    Ben~Davies \altaffilmark{9}
}
 
\email{mmessine@mpifr-bonn.mpg.de}

\altaffiltext{1}{Max-Planck-Institut f\"ur Radioastronomie,
Auf dem H\"ugel 69, D-53121 Bonn, Germany}

\altaffiltext{2}{Department of Physics and Astronomy, The Open University, Walton Hall, Milton Keynes, MK7 6AA, UK 	}

\altaffiltext{3}{Center for Detectors, Rochester Institute of Technology, 54 Memorial Drive, Rochester, NY 14623, USA} 
     
\altaffiltext{4}{Institute for Astronomy, University of Hawaii, 2680 Woodlawn Drive, Honolulu, HI 96822,USA}
     
\altaffiltext{5}{Centro de Astrobiolog\'{\i}a (CSIC-INTA), Ctra. de Torrej\'on a Ajalvir km4, 28850, Torrej\'onde Ardoz, Madrid, Spain}
     
\altaffiltext{6}{Physics and Astronomy Building, 430 Portola Plaza, Box 951547, Department of Physics 
and Astronomy, University of California, Los Angeles, CA 90095-1547, USA}
     
\altaffiltext{7}{European Southern Observatory, Ave. Alonso de C√ördova 3107, Casilcla 19, Santiago, 19001, Chile}
     
\altaffiltext{8}{European Southern Observatory, Karl Schwarzschild-Strasse 2, D-85748 Garching bei Munchen, Germany}
    
\altaffiltext{9}{Astrophysics Research Institute, Liverpool John Moores University,
Twelve Quays House, Egerton Wharf, Birkenhead, Wirral.
CH41 1LD, United Kingdom. }   	   

\altaffiltext{10}{Max-Planck-Institute for Astrophysics, Karl-Schwarzschild-Str. 1, 85748 Garching, Germany}

\altaffiltext{11}{European Space Agency (ESA), The Astrophysics and Fundamental Physics Missions Division, 
Research and Scientific Support Department, Directorate of Science and Robotic Exploration, 
ESTEC, Postbus 299, 2200 AG Noordwijk, The Netherlands}

\begin{abstract}
Rich in HII regions, giant molecular clouds  are natural laboratories to
study  massive stars and sequential star formation.  
The Galactic star forming complex W33 is located at $l= \sim 12\fdg8$ and at a distance of 2.4 kpc,
has a size of $\approx 10$ pc and a total mass of $\approx (0.8 -− 8.0) \times 10^5$ \Msun.
The integrated  radio and IR luminosity of W33 - when combined with the direct detection of methanol masers,
 the protostellar object W33A, and protocluster embedded within the radio source W33 main - mark the region out as
a site of vigorous ongoing star formation. In order to assess the long term 
star formation history, we performed an infrared spectroscopic search for massive stars, 
detecting for the first time fourteen early-type stars, including one WN6 star and four O4-7 stars. 
The distribution of spectral types suggests that this population formed during the last $\sim2-4$ Myr, 
while the absence of red supergiants  precludes extensive star formation at ages 
$6-30$ Myr.  This activity appears distributed throughout the region and does not appear to have yielded the  
dense stellar clusters that characterize other star forming complexes such as Carina and G305. Instead,
we anticipate that W33 will eventually evolve into a loose stellar aggregate, with Cyg OB2 serving as  
a useful, albeit richer and more massive,  comparator. Given recent distance estimates, and despite 
a remarkably similar stellar population,  the rich cluster Cl 1813$-$178 located on the north-west edge
of W33 does not appear to be physically associated with W33.

\end{abstract}





\section{Introduction}    

Massive stars enrich the galactic interstellar medium via the feedback of radiative and mechanical energy, 
the deposition of chemically processed gas via their strong winds and, latterly, solid state material during 
the post main-sequence (MS) phase. Because of their luminosities, individual massive stars  can be detected 
and resolved   in external galaxies, providing direct measures of distances and spatially resolved  metallicity 
gradients  \citep[e.g.,][]{kudritzki14}. At and beyond the end of their lives they power a wide variety of highly energetic transient 
phenomena - firstly during their deaths in supernovae or gamma-ray bursts 
and subsequently by accretion onto their stellar corpses in X-ray binaries 
\citep[e.g.,][]{gudel09,eldridge13}.

Considerable uncertainty remains regarding the mechanism(s) of formation  of massive stars, 
although it is strongly suspected that this process is hierarchical: massive stars 
are found in  apparently isolated young massive stellar clusters 
\citep[e.g., the Arches and Quintuplet clusters;][]{figer99}, in loose 
associations \citep[e.g., Cyg OB2;][and refs. therein]{wright14,negueruela08},
and in large molecular complexes \citep[e.g., 30 Doradus and G305;][]{walborn97,clark04}.

Massive stars are very often part of  binary systems 
\citep[ typically, a fraction of 91\% of OB stars is found
to have  companions, ][]{sana14}.  
A population of apparently isolated massive stars also exists, although it is not clear whether these have 
genuinely formed in isolation  \citep[e.g.,][]{bestenlehner11} or, instead, were lost from a natal aggregate 
due to  dynamical or supernova driven ejection \citep[runaway stars,][]{oh14, povich08}.

Because of the high and variable interstellar extinction and uncertain distances of stars within the 
Galactic Disk, it has long been suspected that our census of massive star 
forming regions is incomplete. Fortunately, the plethora of modern infrared and radio surveys - 
 e.g., MAGPIS, GLIMPSE, WISE, MSX, 2MASS, UKIDSS, and VVV
\footnote{MAGPIS stands for The Multi-Array Galactic Plane Imaging Survey 
\citep[][]{white05,helfand06}, 
2MASS for Two Micron All Sky Survey   \citep{cutri03}, 
DENIS for Deep Near Infrared Survey of the Southern Sky  \citep{epchtein94}, 
UKIDSS for UKIRT Infrared Deep Sky Survey   \citep[][]{lucas08},
VVV for the VISTA Variables in the Via Lactea  survey  \citep{soto13}, 
MSX for Midcourse Space Experiment (MSX) \citep{egan03,price01}, 
GLIMPSE for  Galactic Legacy Infrared Mid-Plane Survey Extraordinaire  \citep{spitzer09}, 
and WISE for Wide-field Infrared Survey Explore \citep{wise12}.},
 - allow us to identify both the natal giant molecular clouds (GMCs) and the stars that form within them. 
Subsequent analysis of the physical properties of  GMCs and associated stellar population(s) -
in terms of the mass function of pre-stellar clumps/cores (proto-stars),
and already formed massive stars, and their temporal and spatial distributions - 
enable constraints to be placed on the mode of star formation that occurred in the region in question
\citep[e.g.,][]{messineo14a}.

One such massive star forming region is the W33 complex, located in  the Galactic plane at
longitude $l= \sim 12\fdg8$; a parallactic distance of $2.40^{+0.17}_{-0.15}$ kpc was 
determined from observations of water masers, which  suggests a location in the Scutum 
spiral arm \citep{immer13}. Subtending 15\arcmin\ ($\sim10$ pc), 
it comprises a number of distinct molecular and/or dusty condensations 
(see Immer et al. 2014 for a census), 
with an integrated IR luminosity of $\sim8\times10^5$ \Lsun\
and a total mass of $\sim(0.8-8)\times10^5M_{\odot}$ \citep{immer13}. Radio observations of one component - 
W33 Main - revealed the presence of an obscured (proto-)cluster apparently comprising a number of stars 
with spectral types ranging from O7.5 to B1.5 \citep[][]{haschick83}.  
The presence of ongoing massive star formation is also signposted by the 
presence of OH, H$_2$O, and CH$_3$OH masers \citep{immer13}, and the direct identification of a bipolar outflow 
and massive dusty torus associated with the  young stellar object W33A \citep{davies10}.

Independently of these studies, Messineo et al. (2008, 2011) serendipitously identified a hitherto overlooked 
young massive cluster - Cl 1813$-$178 - in the vicinity of W33. Analysis of the post-MS content of the cluster 
suggested a mass $\gtrsim10^4M_{\odot}$, making it amongst the most massive aggregates in the Galaxy 
\citep{clark13}.  Given the unusual mix of spectral types present,  \citet{messineo11} quoted 4-4.5 Myr, 
but highlighted that  several cluster members had low luminosities for that age; 
stellar luminosities would appear to demonstrate some degree of non-coevality.
Intriguingly, Cl 1813$-$178 is found  in the vicinity of the pulsar wind nebula  
HESS J1813$-$178 \citep{helfand07,messineo08}. 
 With a spin-down measurement of  44.7ms and a spin-down  luminosity of 
$\dot{E}\sim5.6\times10^{37}$erg s$^{-1}$, 
PSR J1813$-$1749 is one of the youngest and most energetic pulsars in the Galaxy \citep{halpern12}. 
While the energetic young pulsar potentially 
lies beyond both regions \citep[$>4.8$ kpc,][]{halpern12}; 
evidently, this line-of-sight samples numerous regions 
of massive star formation.

Here, we present a near-infrared spectroscopic survey of bright stars
in selected regions of W33  \citep[cl1, cl2, and Mercer1,][]{messineo11},  
and  in the nearby GLIMPSE bubble N10 \citep[e.g.,][]{churchwell06}.
We aim to determining the massive stellar content of W33, its  star formation history, 
and, hence,  relation to the nearby cluster Cl 1813$-$178 and pulsar PSR J1813$-$1749.
In Sect.\ \ref{observations}, we present the spectroscopic observations, and in Sect.\
\ref{photometry}, the available infrared photometry.
In Sect.\ \ref{observations}, we present the spectroscopic observations, and in Sect.\
\ref{photometry}, the available infrared photometry.
In Sect.\ \ref{analysis}, we describe the spectral features and stellar properties.
In Sects.\ \ref{w33main},  \ref{N10}, and \ref{cl1813},  we briefly describe the 
spatial and temporal distributions of the detected stars, before summarising our
findings in Sect. \ref{summary}.

\section{Targets and spectroscopic observations }
\label{observations}

\begin{figure*}
\begin{center}
\resizebox{0.4\hsize}{!}{\includegraphics[angle=0]{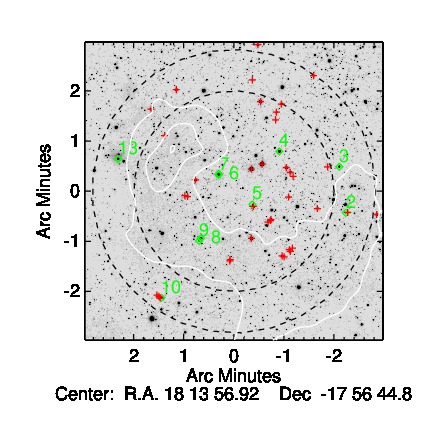}}
\resizebox{0.4\hsize}{!}{\includegraphics[angle=0]{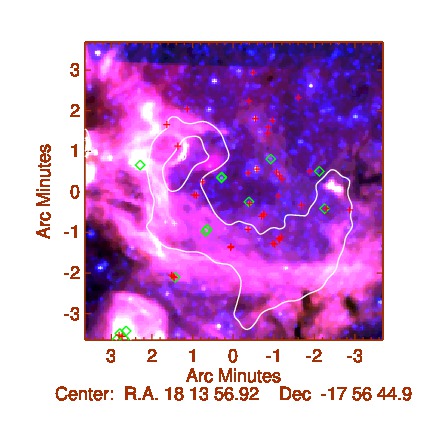}}
\end{center}
\begin{center}
\resizebox{0.4\hsize}{!}{\includegraphics[angle=0]{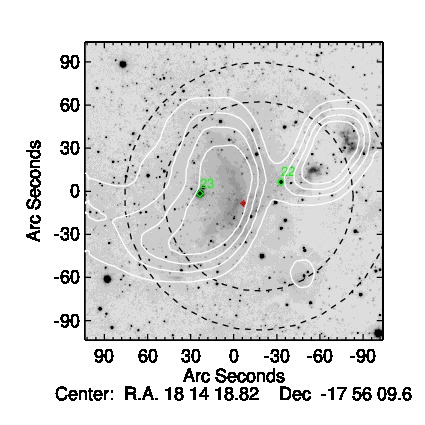}}
\resizebox{0.4\hsize}{!}{\includegraphics[angle=0]{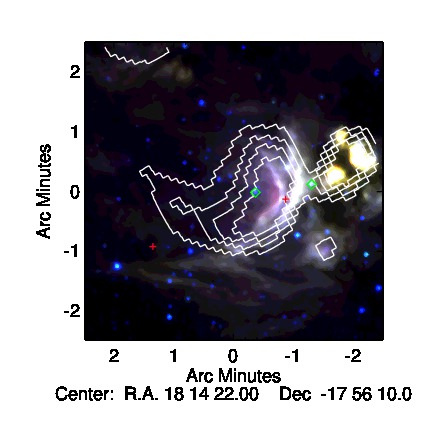}}
\end{center}
\begin{center}
\resizebox{0.4\hsize}{!}{\includegraphics[angle=0]{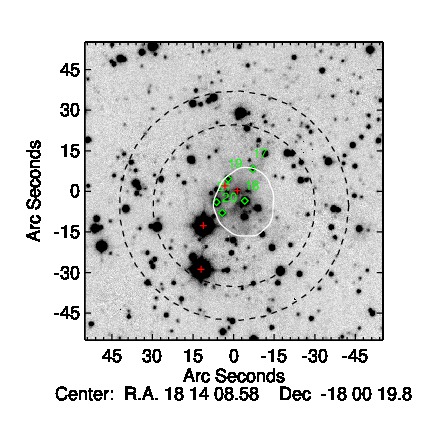}}
\resizebox{0.4\hsize}{!}{\includegraphics[angle=0]{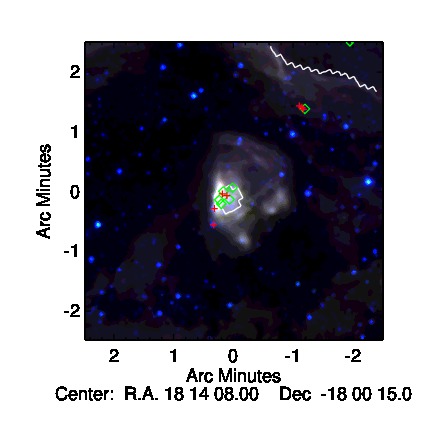}}
\end{center}
\caption{\label{strange} 
Images of the stellar clusters Mercer1 ( top row), 
cl1 ( middle row), and cl2 ( bottom row) 
in W33 \citep{messineo11}.
 Left side: detected stars are displayed on  a 
UKIDSS $K$-band image (for cl2 the $J$ image is used).
Diamonds and crosses indicate the positions of detected early-types and late-types, 
respectively.  Contours of  the 90 cm radio-continuum 
emission from 100 up to 400 mJy beam$^{-1}$ (with a step 
of 100 mJy beam$^{-1}$) are shown in white. 
The two dashed circles mark the regions used for the CMDs in Fig.\ \ref{CMDs}.
By assuming a distance of 2.4 kpc, one arcminute corresponds to 0.70 pc.
 Right side:  contours of  the 90 cm radio-continuum 
emission superimposed on a GLIMPSE composite image 
(3.6 \um\ in blue, 5.8 \um\ in green, and 8.0 \um\ in red). 
} 
\end{figure*}

Spectroscopic targets  with 2MASS \Ks\ from 6 mag to 11 mag  and 
$H-$\Ks $> 0.5$ mag were  selected from the regions listed in \citet{messineo11}
that exhibited  over-densities of bright stars or 
nebulae in 2MASS and GLIMPSE images.
The Mercer1 region\footnote{\citet{messineo11} define this as a region with 
a radius of 2\farcm3  that includes
the candidate stellar cluster n. 1 of \citet{mercer05}.}, to the west of the radio source W33 Main,
appears as a sparse aggregate of bright stars with a pronounced arc of IR and radio 
emission to the south that is suggestive of a wind blown structure (Fig.1, top panel). 
The cl1 region, immediately to the east of the embedded proto-cluster in W33 Main, 
contains an isolated bright star surrounded by an  arc of emission at IR and radio wavelengths 
(Fig. \ref{strange}, middle panel). The cl2 region, coincident with W33 B1 
\citep[e.g.,][]{immer14}, contains a 
 small cluster of stars associated with diffuse IR and radio emission 
 (Fig. \ref{strange}, bottom panel).
These targets were supplemented with a few isolated bright targets selected on the basis of  their 
GLIMPSE colors - e.g., indicative of free-free emitters
\citep[e.g.,][]{hadfield07,messineo12}) - such as star \#1 and
the candidate luminous blue variable (LBV) \#15 from Cl 1813$-$178.
Additionally, we observed some stars to the north of W33, 
in the cluster  BDS2003-115 embedded in the  GLIMPSE mid-IR
bubble N10 (Fig. 2).

Data were acquired with the Spectrograph for INtegral Field Observations 
in the Near Infrared   \citep[SINFONI,][]{eisenhauer03} on  the Yepun Very 
Large Telescope, under the ESO programs 087.D-0265(A) and 089.D-0790(A).
The $K$-grating was used along with the $0\rlap{.}^{\prime\prime}25$ pix$^{-1}$ 
scale to yield a resolving power of $\sim4500$.
Integration times per exposure ranged from 30 s to 300 s. 
Typically, each observation consisted of four exposures, two on target and two on sky.
Data-cubes  were generated 
with the version 3.9.0 of the ESO SINFONI  pipeline  
\citep{schreiber04,modigliani07},  using  flat-fields,
bad-pixel masks, distortion maps, and  arcs.
From each cube, stellar traces with signal-to-noise ratio larger than 30-40 
were analyzed. Corrections for instrumental and atmospheric responses
were accomplished with standard stars of B-types; stellar \brg\  and  
He I lines at 2.112 \um\ were removed from the standard 
spectra with a linear interpolation.
A sky subtraction was performed to eliminate possible nebular lines and 
residuals from OH subtraction. 
A total of 86 cubes were observed  and 94 stars were extracted.

\begin{figure*}
\begin{center}
\resizebox{0.4\hsize}{!}{\includegraphics[angle=0]{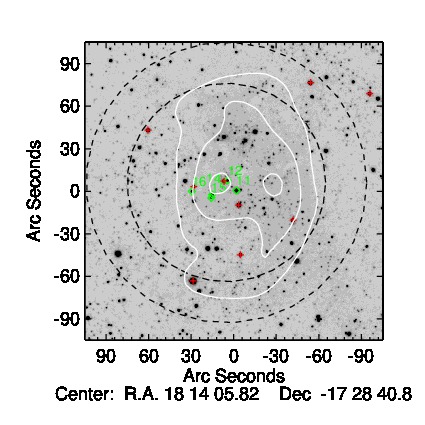}}
\resizebox{0.4\hsize}{!}{\includegraphics[angle=0]{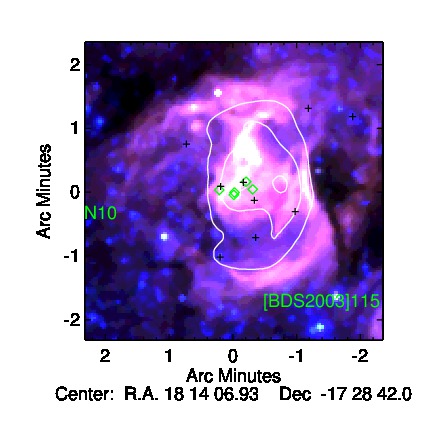}}
\end{center}
\caption{\label{strange2} 
Images of the bubble N10.
{\it Left side:} Detected stars  are displayed on  a UKIDSS $K$-band image.
Diamonds and crosses indicate the positions of detected early-types and late-types, 
respectively.
The white contours show  the 90 cm radio-continuum 
emission of the bubble N10 from MAGPIS  at 100, 200, and 300 mJy beam$^{-1}$.
By assuming a distance of 4.29 kpc, one arcminute corresponds to 1.25 pc.
{\it Right side: } 
The 90 cm contours are superimposed on a GLIMPSE composite image 
(3.6 \um\ in blue, 5.8 \um\ in green, and 8.0 \um\ in red). }
\end{figure*}

\section{Available  photometric data}
\label{photometry}
We searched for counterparts of the spectroscopically observed stars
in the 2MASS Catalog of Point Sources \citep{cutri03}, in
the UKIDSS catalog  \citep[UKIDSS,][]{lucas08},
and in the DENIS catalog   \citep[DENIS,][]{denis05} 
by taking the closest match within 1\arcsec.
Mid-infrared data were retrieved from  
the MSX  \citep[MSX,][]{egan03} with a search radius of 5\arcsec,
from the  GLIMPSE 
\citep[GLIMPSE,][]{spitzer09} and   The WISE
\citep[WISE,][]{wise12} catalogues with a search radius of 2\arcsec.

For 21\% of the sources, $R$-band counterparts with magnitudes from 12.38 mag 
to 19.80 mag were found in the  
The Naval Observatory Merged Astrometric Dataset (NOMAD) \citep{zacharias04}.

Photometric measurements  are listed  in the appendix.
For the observed targets, there is no additional information 
from the SIMBAD database.

\subsection{UKIDSS photometry}
For stars fainter than \Ks$\sim 10.5$ mag, we used UKIDSS photometry  \citep{lucas08}.
$JHK$ magnitudes  are available from the UKIDSS data release number 6 
(DR6) \citep{lucas08}; however,  for four fields (Mercer1, cl1, cl2, N10) 
we generated photometric catalogs 
of point sources   with  the psf-fitting algorithm DAOPHOT  
\citep{stetson87} and the  leavestack frames provided by UKIDSS \citep{lucas08}. 
A detection threshold of   4 $\sigma$ was used.
From  overlapping  fields, independently calibrated with 2MASS datapoints, 
an absolute photometric error of 0.05 mag was estimated. 
Comparison of the UKIDSS pipeline  and the psf-fitting catalogs in the 
Mercer1 field  yielded an absolute uncertainty of 0.07 mag.
The resultant  magnitudes are given in the appendix.

\section{Analysis}
\label{analysis}

\subsection{Spectral classification}

We detected a total of 23 new early-types,  which 
comprise  one Wolf-Rayet, 13  stars of  spectral type O or B, 
 and  nine stars with indeterminate, but early,  spectral types  (OBAF),
and 70 late-type stars (see Tables \ref{table.obspectra}, 
\ref{otypetable}, and \ref{table:obslate}).

 \begin{table}
 \begin{center}
\caption{\label{table.obspectra} Summary of Observations of Early-type Stars.}
\begin{tabular}{ r c c r r r r r r}
\hline
\hline
ID      &  Ra[J2000]  & Dec[J2000]      &   Spec.             &    Period           &   Date-Obs   \\
        &  {\rm [hh mm ss]}   & {\rm [deg mm ss]} &          &           &    yyyy-mm-dd            \\ 
\hline
  M15$^a$    & 18 13 20.99 & -17~49~46.9     &            cLBV     &   P89     &          2012-06-21    \\
  1     & 18 13 34.81 & $-$18~05~41.5     &             WN6     &             P89     &                2012-06-21    \\
  2     & 18 13 47.53 & $-$17~57~10.7     &            OBAF     &             P89     &                2012-06-27    \\
  3     & 18 13 48.07 & $-$17~56~15.6     &            B0-5     &             P87     &                2011-08-19    \\
  4     & 18 13 53.10 & $-$17~55~57.3     &            B0-5     &             P87     &                2011-05-29    \\
  5     & 18 13 55.36 & $-$17~57~00.7     &            OBAF     &             P87     &                2011-05-29    \\
  6     & 18 13 58.19 & $-$17~56~23.9     &            B0-5     &             P87     &                2011-05-29    \\
  7     & 18 13 58.20 & $-$17~56~25.4     &            O4-6     &             P87     &                2011-05-29    \\
  8     & 18 13 59.69 & $-$17~57~41.2     &            O4-6     &             P87     &                2011-08-19    \\
  9     & 18 13 59.83 & $-$17~57~44.4     &            OBAF     &             P87     &                2011-08-20    \\
 10     & 18 14 03.00 & $-$17~58~52.3     &            B0-5     &             P87     &                2011-08-20    \\
 11     & 18 14 05.69 & $-$17~28~40.3     &            O4-6     &             P87     &                2011-06-24    \\
 12     & 18 14 06.12 & $-$17~28~33.2     &            OBAF     &             P87     &                2011-06-24    \\
 13     & 18 14 06.63 & $-$17~56~06.3     &            B0-5     &             P87     &                2011-08-26    \\
 14     & 18 14 06.89 & $-$17~28~43.2     &             OBe     &             P89     &                2012-09-15    \\
 15     & 18 14 06.93 & $-$17~28~45.5     &            OBAF     &             P89     &                2012-09-15    \\
 16     & 18 14 07.89 & $-$17~28~40.9     &            OBAF     &             P89     &                2012-09-15    \\
 17     & 18 14 08.09 & $-$18~00~11.6     &            B0-5     &             P89     &                2012-06-26    \\
 18     & 18 14 08.30 & $-$18~00~23.3     &            B0-5     &             P87     &                2011-05-29    \\
 19     & 18 14 08.74 & $-$18~00~15.2     &            OBAF     &             P89     &                2012-06-26    \\
 20     & 18 14 08.89 & $-$18~00~27.8     &            OBAF     &             P89     &                2012-09-15    \\
 21     & 18 14 09.03 & $-$18~00~23.8     &            OBAF     &             P89     &                2012-09-15    \\
 22     & 18 14 16.52 & $-$17~56~03.2     &              Oe     &             P87     &                2011-05-29    \\
 23     & 18 14 20.48 & $-$17~56~11.2     &            O6-7     &             P87     &                2011-05-29    \\
\hline
\end{tabular}
\begin{list}{}{}
\item[$^a$] Star $[$MFD2011$]$ 15 was discovered by \citet{messineo11}.  
\end{list}
\end{center}
\end{table}

\subsubsection{Early-type stars}

\begin{figure*}
\resizebox{0.49\hsize}{!}{\includegraphics[angle=0]{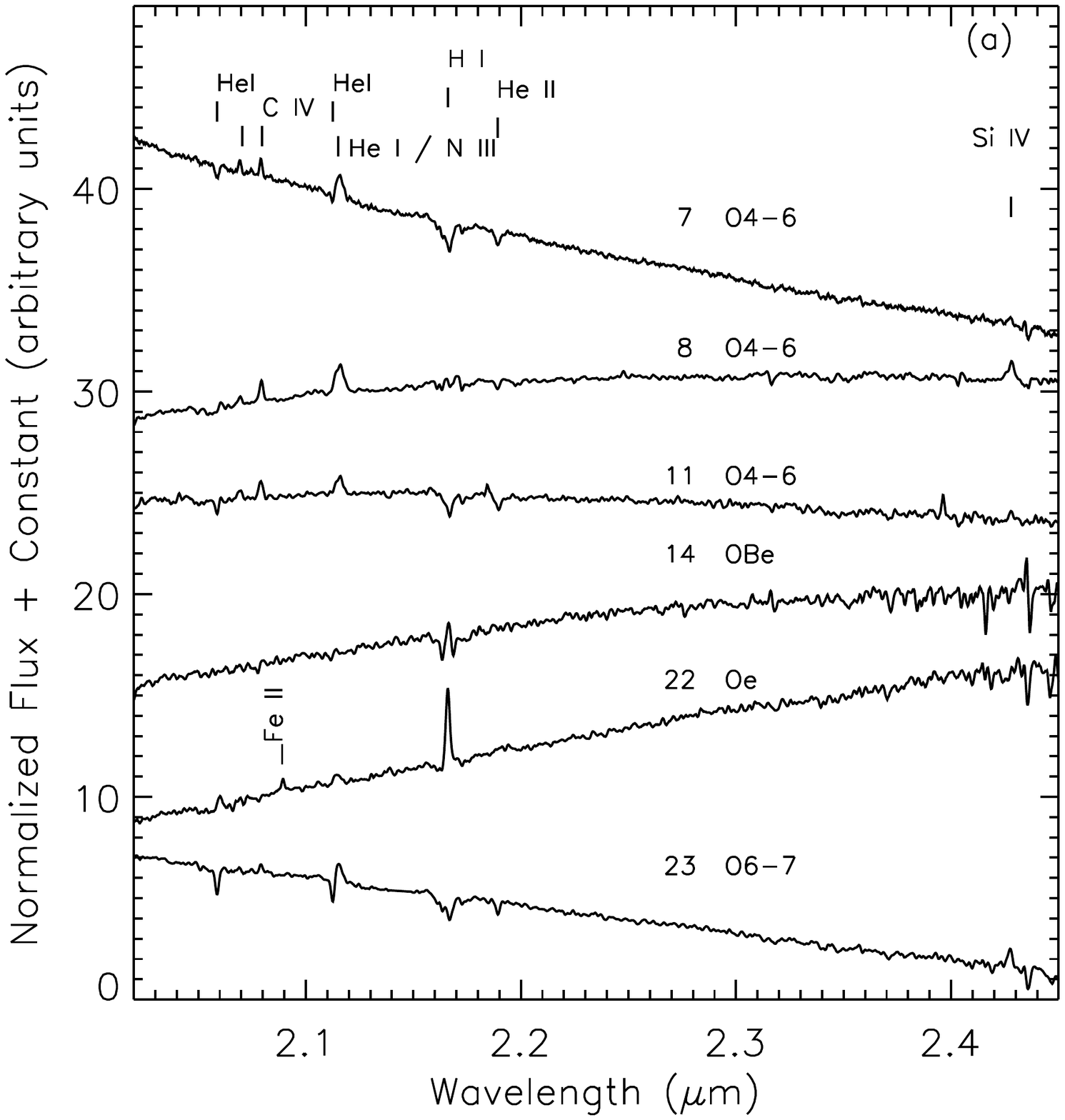}}
\resizebox{0.49\hsize}{!}{\includegraphics[angle=0]{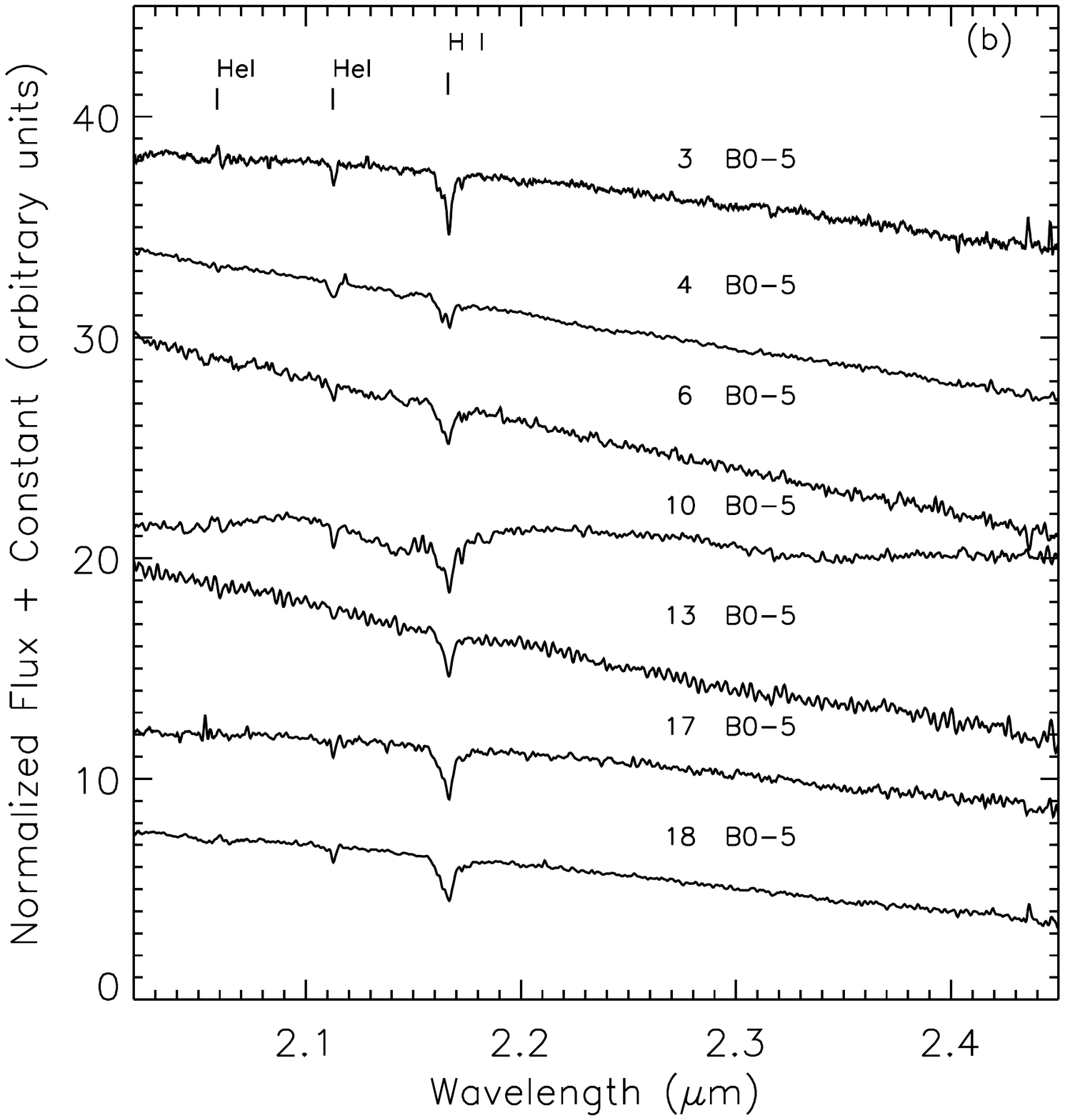}}
\resizebox{0.49\hsize}{!}{\includegraphics[angle=0]{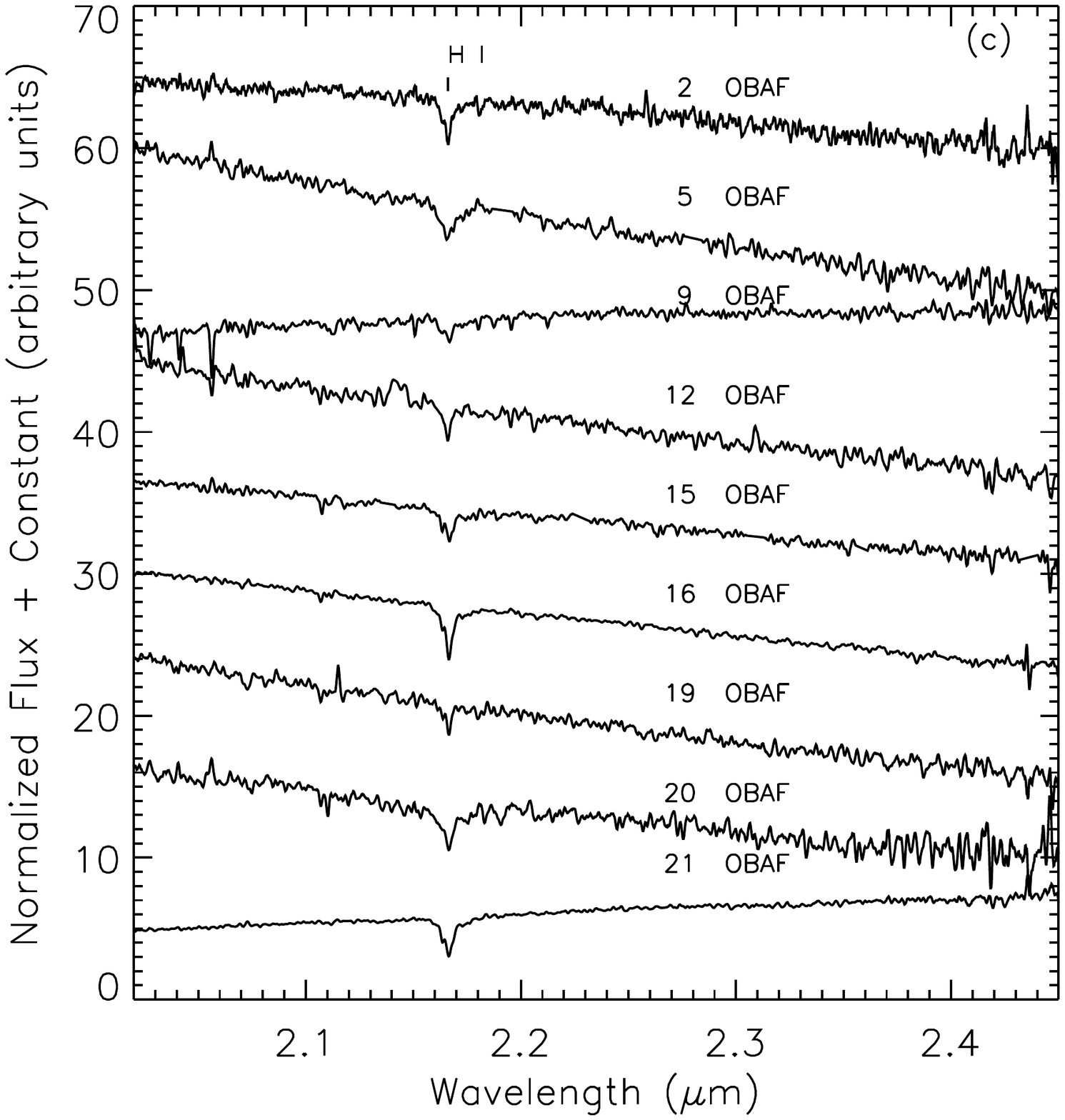}}
\caption{\label{spectra} SINFONI $K-band$ spectra of newly detected O-type  stars.
Panel (a) displays massive O-type stars, panel (b) B-type stars, and panel (c)
spectra (mostly marginal detections) with  a detection of a \brg\ line.} 
\end{figure*}

\begin{figure}
\resizebox{1.0\hsize}{!}{\includegraphics[angle=0]{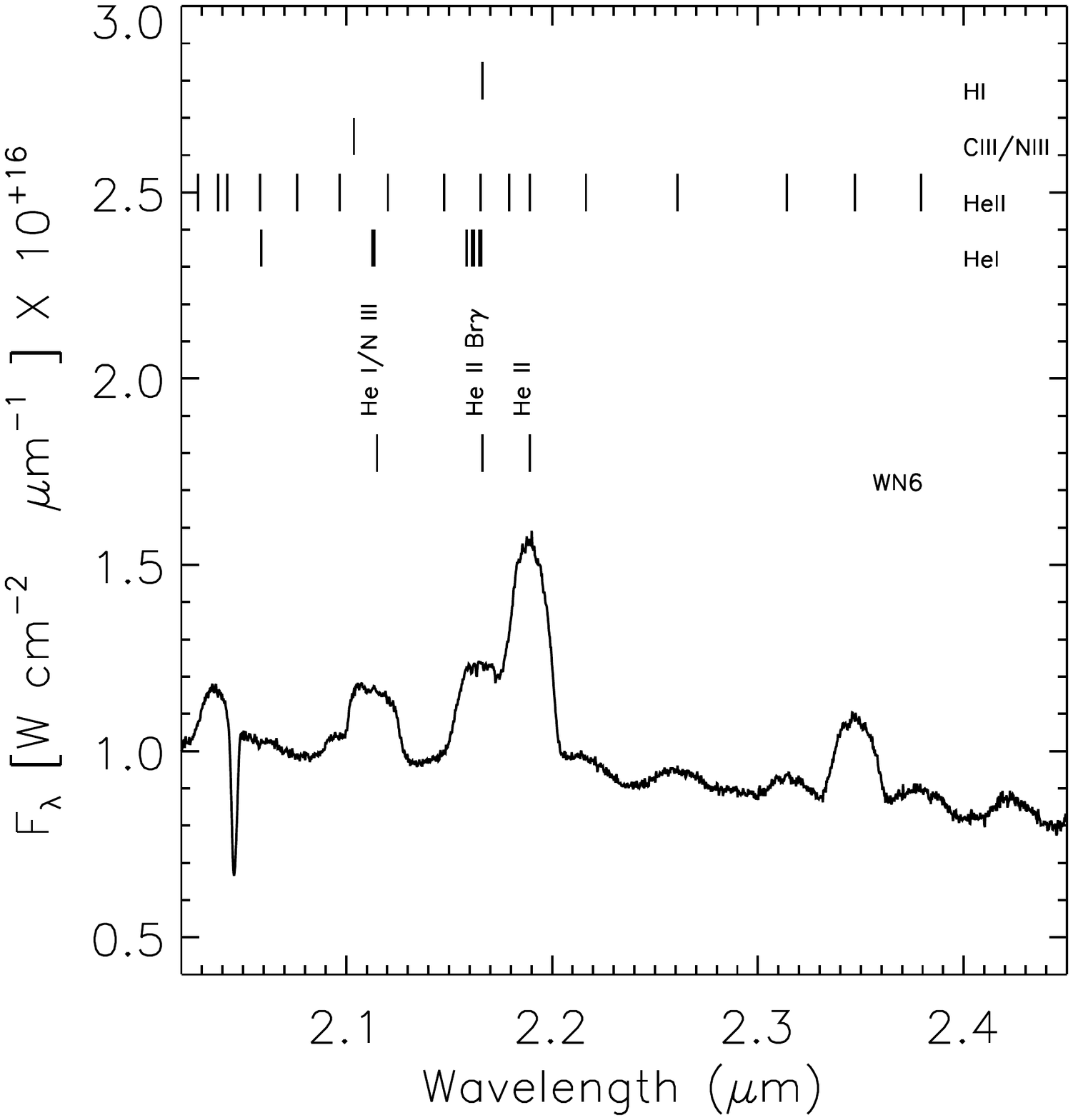}}
\resizebox{1.0\hsize}{!}{\includegraphics[angle=0]{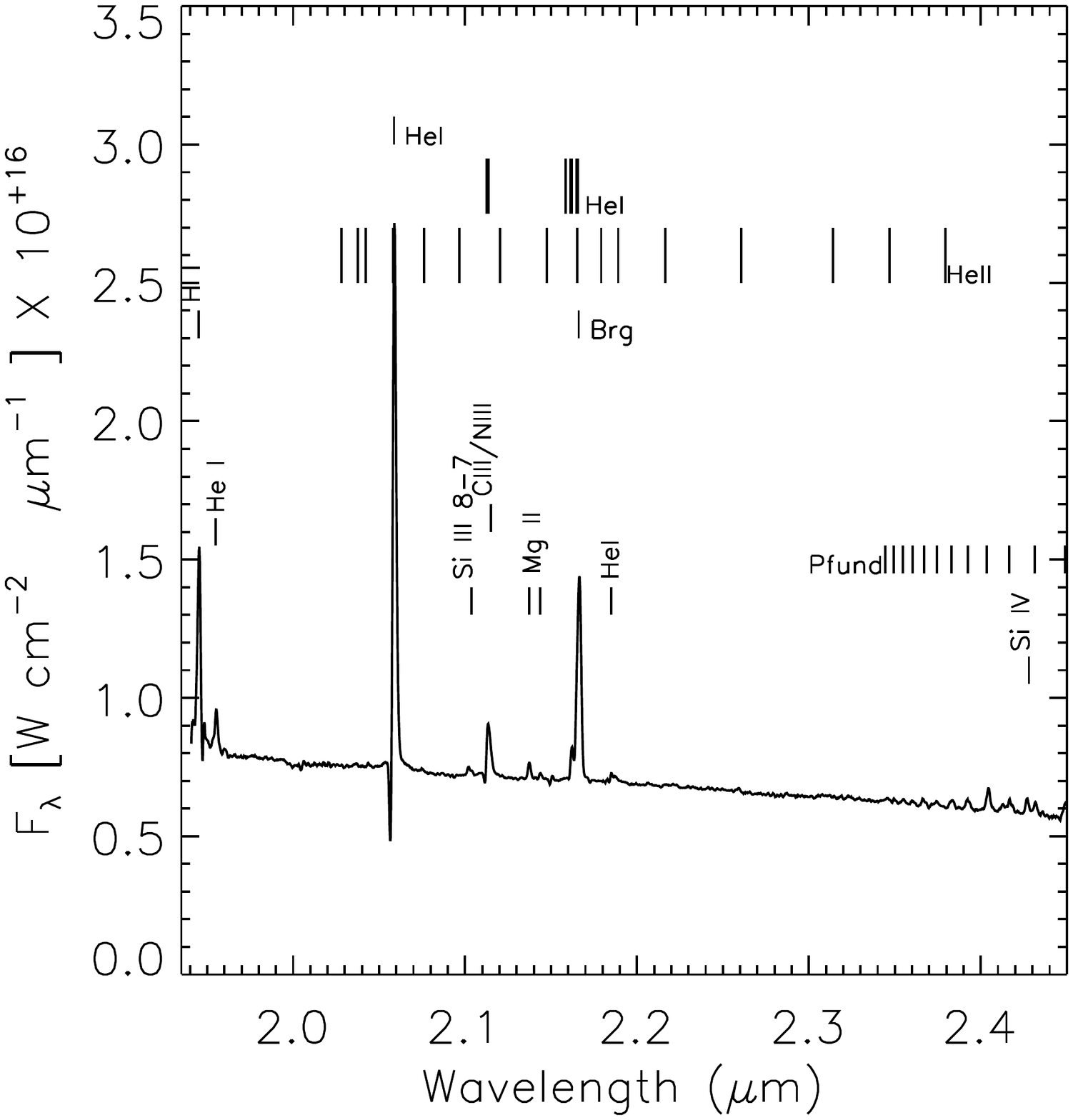}}
\caption{  \label{newWR.fig} \label{cLBVfig} 
{\it Upper panel:} SINFONI $K$-band spectrum of a newly detected WN6 star. 
{\it Lower panel:} SINFONI $K$-band spectrum of the candidate LBV  [MFD2011] 15.} 
\end{figure}

The spectra  of the newly-discovered early-type stars are  characterized by 
hydrogen and helium  lines, as well as transitions of 
heavier elements, such as C, N, O, and Fe. Their spectral classification  
was accomplished by comparison to the
near-infrared atlases of
\citet{hanson96}, \citet{hanson05}, and \citet{figer97}.

Star \#1 is a newly discovered Wolf-Rayet star; its spectrum is characterized 
by   strong and broad emission lines;  He I/ N III  centered at 2.1117 \um, 
 He II/ \brg\ line at 2.1636 \um, and He II at 2.1891 \um. 
The equivalent width ratios between  the 2.1891 \um\ line
and the two lines at 2.1636 \um\ and 2.112 \um\ were estimated  to be $1.8\pm0.3$ and 
$4.3\pm0.7$ \footnote{ Errors on the ratios are calculated by propagating the errors
on the EWs; for each line, errors on the EWs are obtained with the formula
of \citet{vollmann06}.  
For the \brg\ and He II line at 2.189 \um, EWs are calculated after having 
subtracted from the observed spectrum the gaussian fits to the  contaminating lines.}
respectively, by using  multiple gaussian fits \citep[][]{figer97}.
This spectrum resembles that of WR134,
a WN6b star \citep{figer97};  the suffix b 
indicates  broad emission lines \citep[e.g.,][]{hadfield07};
derived ratios are almost identical to those calculated for  WR134; thereby, 
star \#1 is a WN6b.  


The spectra of stars   \#7, \#8, \#11, and \#23  are characterized by emission in the
C IV  2.0705 \um\ and  2.0796 \um\ and  O III/N III at $\approx 2.115$ \um\ lines,  
He I line at 2.059 \um\ and He II  2.189 \um\ in absorption, and  \brg\   
mostly in absorption (see Table \ref{otypetable} and   Fig. \ref{spectra}). 
This combination of features is characteristic of stars of mid- to late-O spectral type; 
unfortunately, for this temperature range  the assignment of luminosity classes from K-band 
spectra alone is somewhat problematic. Nevertheless, in this regard we note the 
strong morphological resemblance of these objects to the O4-6 I stars within the Arches cluster 
\citep{martins08}. The spectra of stars \#8, \#11, and \#23 display the Si IV emission line 
at 2.428 \um. 
We denote O-type stars with O III/N III and Si IV in emission 
by  Of$_K^{+}$, as described in \citet{messineo14a},  although we caution that 
this does not necessarily indicate a super/hyper-giant classification \citep{dejager98}.
Star \#8 has \brg\ filled in, indicating I+ nature.
The spectrum of star \#23 has the He I line at 2.112 \um\ in absorption, and
most likely has a later sub-type.

\begin{table}
\begin{center}
\caption{\label{otypetable} List of Lines Detected in Early-type Stars}
\begin{tabular}{rrrrrrr}
\hline
\hline
ID  &   Center   & FWHM & EW$^{**}$               &  SN$^+$    & \Vlsr$^{++}$\\
    &    [\um]   & [\AA]&  [\AA]           &      &        [\kms] \\
\hline
M15 &   2.059024 &   19 $\pm$    0 &-52.1 $\pm$   0.4 &  278 &   $63\pm10$  \\
M15 &   2.166333 &   26 $\pm$    0 &-33.6 $\pm$   0.4 &  355 &   $44\pm10$  \\
  1 &   2.045707 &  $33\pm0$       &  12.7 $\pm$  1.0 &    6 &     $..$  \\
  1 &   2.111480 &  $229\pm1$      & $-$43.1 $\pm$  5.8 &   18 &      $..$  \\
  1 &   2.163296 &  $199\pm2$      &$-$106.3 $\pm$ 16.3 &   32 &      $..$  \\
  1 &   2.188899 &  $195\pm1$      &$-$186.2 $\pm$ 14.4 &   68 &     $..$  \\
  2 &   2.165926 &   57 $\pm$    8 &  6.3 $\pm$   6.0 &    7 &     $..$  \\
  3 &   2.113018 &   19 $\pm$    1 &  1.8 $\pm$   0.9 &   10 &     $..$  \\
  3 &   2.166116 &   56 $\pm$    0 &  6.9 $\pm$   1.1 &   27 &     $..$  \\
  4 &   2.112701 &   38 $\pm$    1 &  1.8 $\pm$   0.4 &   12 &     $..$  \\
  4 &   2.165608 &   67 $\pm$    2 &  3.2 $\pm$   1.3 &   10 &     $..$  \\
  $*5$ &   2.165962 &   77 $\pm$  187 &  2.3 $\pm$   7.1&    4 &     $..$  \\
  6 &   2.112678 &   17 $\pm$   11 &  2.0 $\pm$   1.7 &    4 &     $..$  \\
  6 &   2.165574 &   73 $\pm$    8 &  4.1 $\pm$   2.7 &    6 &     $..$  \\
  7 &   2.058728 &   21 $\pm$    4 &  0.7 $\pm$   1.1 &    5 &     $..$  \\
  7 &   2.069326 &   11 $\pm$    4 & -0.2 $\pm$   0.5 &    4 &     $..$  \\
  7 &   2.079126 &   11 $\pm$    1 & -0.5 $\pm$   0.5 &   11 &     $..$  \\
  7 &   2.115881 &   32 $\pm$    0 & -1.5 $\pm$   0.8 &    8 &     $..$  \\
  7 &   2.165977 &   55 $\pm$    2 &  3.5 $\pm$   1.2 &   13 &     $..$  \\
  7 &   2.189267 &   16 $\pm$   14 &  1.9 $\pm$   0.5 &   11 &     $..$  \\
  $*8$&   2.069469 &   13 $\pm$   21 & -0.4 $\pm$   0.7 &    4 &     $..$  \\
  8 &   2.079379 &   16 $\pm$    0 & -1.4 $\pm$   0.6 &   12 &     $..$  \\
  8 &   2.115700 &   46 $\pm$    0 & -4.3 $\pm$   0.7 &   15 &     $..$  \\
  8 &   2.189147 &   15 $\pm$    8 &  0.6 $\pm$   0.9 &    4 &     $..$  \\
  8 &   2.427893 &   38 $\pm$   11 & -2.1 $\pm$   2.0 &    4 &     $..$  \\
  9 &   2.166162 &   65 $\pm$   34 &  3.4 $\pm$   3.9 &    6 &     $..$  \\
 10 &   2.112934 &   17 $\pm$    1 &  0.5 $\pm$   1.3 &    7 &     $..$  \\
 10 &   2.166113 &   92 $\pm$    6 &  6.9 $\pm$   1.8 &   14 &     $..$  \\
 11 &   2.058704 &   13 $\pm$    4 &  0.4 $\pm$   2.1 &    4 &     $..$  \\
 11 &   2.079061 &   20 $\pm$    3 & -1.4 $\pm$   0.8 &    8 &     $..$  \\
 11 &   2.115609 &   38 $\pm$    3 & -2.3 $\pm$   0.7 &   12 &     $..$  \\
 11 &   2.166673 &   34 $\pm$    6 &  2.6 $\pm$   1.5 &   11 &     $..$  \\
 11 &   2.189657 &   20 $\pm$   11 &  1.2 $\pm$   1.1 &    6 &     $..$  \\
 $*11$ &   2.427892 &   17 $\pm$   19 &  0.0 $\pm$   2.7 &    2 &     $..$  \\
 12 &   2.165692 &   57 $\pm$   16 &  8.9 $\pm$   5.1 &    8 &     $..$  \\
 13 &   2.112964 &   21 $\pm$   15 & -0.3 $\pm$   1.1 &    3 &     $..$  \\
 13 &   2.166270 &   54 $\pm$    3 &  5.9 $\pm$   1.4 &   15 &     $..$  \\
 14 &   2.166573 &    $..$         &  1.3 $\pm$   3.6 &    6 &     $..$  \\
 15 &   2.166294 &   81 $\pm$   30 &  7.6 $\pm$   3.9 &    8 &     $..$  \\
 16 &   2.166464 &   61 $\pm$    0 & 10.4 $\pm$   2.4 &   18 &     $..$  \\
 17 &   2.112767 &   10 $\pm$    7 &  0.3 $\pm$   1.2 &    6 &     $..$  \\
 17 &   2.165805 &   78 $\pm$    4 &  5.8 $\pm$   2.1 &   11 &     $..$  \\
 18 &   2.112785 &   19 $\pm$    1 &  0.3 $\pm$   0.4 &   14 &     $..$  \\
 18 &   2.165775 &   84 $\pm$    0 &  5.6 $\pm$   1.4 &   13 &     $..$  \\
 19 &   2.166143 &   44 $\pm$   15 & -0.1 $\pm$   4.6 &    5 &     $..$  \\
 20 &   2.166230 &   92 $\pm$   24 & 11.6 $\pm$   6.9 &    6 &     $..$  \\
 21 &   2.166324 &   71 $\pm$    0 &  8.0 $\pm$   2.2 &   14 &     $..$  \\
 22 &   2.166032 &   16 $\pm$    0 & -5.7 $\pm$   2.5 &   21 &     $..$  \\
 23 &   2.058749 &   16 $\pm$    0 &  0.8 $\pm$   2.0 &    6 &     $..$  \\
 $*23$   &   2.068678 &    7 $\pm$  420 &  0.5 $\pm$   1.1 &    3 &     $..$  \\
 23 &   2.079160 &   16 $\pm$   11 & -0.5 $\pm$   0.5 &    6 &     $..$  \\
 23 &   2.115583 &   25 $\pm$    3 & $..$             &   10 & $..$  \\
 23 &   2.165405 &   96 $\pm$   23 &  3.5 $\pm$   1.4 &   11 &     $..$  \\
 23 &   2.189342 &    8 $\pm$    6 &  1.4 $\pm$   0.6 &   10 &     $..$  \\
 23 &   2.427330 &   27 $\pm$    8 & -1.2 $\pm$   1.5 &    5 &     $..$  \\
\hline
\end{tabular}
\end{center}
\begin{list}{}{}
\item[{\bf Notes.}] ($^{**}$) Errors on the EWs are calculated following 
\citet{vollmann06}.~ ($^+$) SN = flux(peak) / continuum noise.
~($*$) = Marked ID numbers indicate hints for lines (with a peak SN$\ga 2$) 
with poor measurements. ~($^{++}$) The used SINFONI setting allows for an
absolute wave calibration within 10 \kms. For star M15, the average offset of detected OH lines 
from their rest wavelengths is 3 \kms, with  $\sigma=8$ \kms. Quoted errors
are sqrt(centererr$^2$+$10^2$).
\end{list}
\end{table}

The spectrum of star  \#22  is characterized by emission in  He I 2.059 \um\  (weak), 
Fe II  2.08958 \um, probable N III 2.11467 \um, and   \brg\ (strong).
O III / N III is a signature of massive  stars from O2 to O8; usually, O4-O7 stars
have additional C IV lines, although they are faint in dwarfs. 
Star \#22  appears to still be partially enshrouded;  
the iron emission, which is indicative of shocks,
is located at  the  position of the star, with  
diffuse H$_2$ emission in its surroundings \citep[lines 1-0 (S1), 1-0 (S0), 2-1 (S1), 
1-0 (Q1), 1-0 (Q2), and 1-0 (Q3)][]{black88,gautier76,scoville83}.

The spectrum of star \#14 shows only a \brg\ line in absorption with a central emission peak. 
Stars \#14 and \#22 have spectral morphologies that are reminiscent of 
stars associated with ultra compact HII regions \citep{bik05,bik06}
and, pre-empting Sect. 4.2, IR excesses that are suggestive 
of emission  from  natal 
circumstellar envelopes. We, therefore, conclude that both
are likely to be very young, early-type stars. 

The spectra of stars \#3, \#4, \#6, \#10, \#13,  \#17, and  
\#18  show both He I 2.112 \um\ and \brg\ 
in absorption, indicative of spectral types  B0 to B5.
The spectra of stars \#2, \#5, \#9, \#12, \#15, \#16, \#19, \#20, and \#21 
are noisy, however, a \brg\ line in absorption is clearly visible. 
They have spectral-types earlier than G-types.

\subsubsection{The candidate LBV [MFD2011] 15}

During the spectroscopic campaign, we  re-observed
[MFD2011] 15, the luminous blue variable (LBV) candidate \#15
in the cluster Cl 1813$-$178 \citep{messineo11} 
-- in order to search for the spectroscopic variability characteristic of this phase of stellar 
evolution. 

[MFD2011] 15 was first identified with NIRPEC observations \citep{mclean98} with R=1900  
by \citet[][]{messineo11}; our new SINFONI spectrum benefits from twice the spectral resolution 
and an improved signal to noise ratio and is shown in Fig. \ref{cLBVfig}. Comparison to the earlier 
spectrum shows an almost identical morphology, with a strong P-Cygni line in He I  at 2.059 \um\ and, 
as well as single peaked emission in He I/ N III/ C III at 2.11407 \um, Mg II at  2.13764 \um\ and 
2.14411 \um,  \brg\ at 2.16655 \um, and He I at 2.185 \um. 
The larger baseline of the SINFONI detector and higher resolving power 
allow detections of H I at 1.94552 \um, He I at 1.95556 \um,  a line emission at 2.10224 \um\ 
(likely due to  Si III 8-7), and a forest of Pfund H lines; 
intriguingly, we also detected  Si IV at 2.42724 \um.
With the exception of the Si IV  transition the spectrum of [MFD2011] 15 bears a strongly 
resemblance to that of the {\em bona fide} LBV P Cygni \citep[e.g.,][]{clark11}.  
The lack of He II lines results in a degeneracy on the temperature estimate
\citep{najarro94, najarro97, martins07}. Nevertheless, the presence of
Si IV  in emission may point to a higher temperature than previously assumed from quantitative analysis 
($\sim16$kK; Messineo et al. 2011), and, therefore, higher luminosity;
this is of interest since the current estimate 
(log$(L/L_{\odot}\sim5.3$) places the star amongst the faintest of 
known (candidate) LBVs  \citep{clark09b}. 
[MFD2011] 15 is included in a sample of Galactic LBVs that is homogeneously remodeled
(Del Mar Rubio-Diez et al. in preparation).

\subsubsection{Late-type stars}
$K$-band spectra of late-type stars are characterized by CO bands in absorption with
a band-head at 2.2935 \um.
We corrected these spectra for interstellar extinction by using
the extinction law by \citet{messineo05}  and the $J,H$ and \Ks\ magnitudes given in Table 6.
We measured  the CO equivalent widths from 2.290 to 2.320 \um\ 
with a continuum from 2.285 to 2.290 \um, as in \citet{figer06}.
The initial assumption of an  average intrinsic $J-$\Ks=1.05 mag
($H-$\Ks=0.23 mag) introduces an uncertainty in \Aks\ of only 0.075 mag, 
which is negligible for spectral classification. A shift by 10\% on the
$K$-band extinction produces a median  shift  in EW 
of 1.4\%. We estimated the uncertainty due to the continuum region adopted 
by adding small shifts and remeasuring; the percentage uncertainty has a 
median value of 5\%. The list of detected late-types is provide in 
Table \ref{table:obslate}.

Spectral types were obtained by comparison with template spectra of  red giants 
and red supergiants  \citep{kleinmann86}. For each star, spectral types 
are provided  for  two possible luminosity classes - giants and supergiants - 
which follow differing relations between  EW(CO)s and spectral-types \citep[e.g.,][]{figer06,messineo14a}.
Typically, the  spectral types obtained via this methodology are accurate to within two spectral types.
For thirteen stars in Table \ref{table:obslate} we found EW(CO)s larger than 52 \AA, i.e., 
larger than those of a M7 III; the spectra of seven of them (\#30, \#47, \#50, \#54, \#57, \#87,  and \#89) 
show  water absorption, and are likely Mira-AGB stars  \citep[see, e.g.,][]{messineo14zhu};
the spectra of stars \#51, \#53, and \#85 are quite noisy; the spectra of
stars \#59, \#61, and \#62  do not show water absorption  and could be
semi-regular asymptotic giant branch stars \citep{messineo14zhu}.

\begin{table*}
\begin{center}
\caption{ \label{table:obslate} List of Detected $G$, $K$, $M$-type Giants.}
\begin{tabular}{@{\extracolsep{-.04in}} rccrrrr|rccrrrrr}
\hline
\hline
ID  & Ra[J2000] & Dec[J2000] &  EW(CO)           & Sp$^*$        & Sp$^*$     &  Obs. Date &  ID  & Ra[J2000] & Dec[J2000] &  EW(CO)           & Sp$^*$     &    Sp$^*$     &  Obs. Date &  \\
    &           &            &                   &   RSG     & RGB    &            &      &           &            &                   &RSG     &  RGB      &            &       \\
\hline
 24 & 18 14 18.4& $-$17 56 18&   22.4$\pm$    5.4&   $<$K1&   K1&          2011-05-29&     59 & 18 13 53.8& $-$17 57 19&   64.9$\pm$    3.1&   M3& $..$&          2011-05-29\\
 25 & 18 13 59.1& $-$17 27 31&   48.6$\pm$    1.6&   K5&   M7&          2011-08-12& 60 & 18 13 53.8& $-$17 57 18&   21.3$\pm$    1.9&   $<$K1&   K1&          2011-05-29\\
 26 & 18 14 07.8& $-$17 29 44&   32.3$\pm$    1.7&   K1&   M0&          2011-08-19& 61 & 18 13 55.3& $-$17 57 03&   57.2$\pm$    1.2&   M1& $..$&          2011-05-29\\
 27 & 18 14 02.0& $-$17 27 24&   26.0$\pm$    1.8&   $<$K1&   K3&          2011-08-20& 62 & 18 13 55.4& $-$17 57 41&   52.9$\pm$    1.9&   M0& $..$&          2011-05-29\\
 28 & 18 10 52.0& $-$17 42 23&   44.5$\pm$    1.1&   K4&   M5&          2011-06-24& 63 & 18 13 52.3& $-$17 56 51&   48.4$\pm$    1.4&   K5&   M7&          2011-05-29\\
 29 & 18 10 51.8& $-$17 42 20&   25.9$\pm$   19.8&   $<$K1&   K4&          2011-06-24& 64 & 18 13 52.2& $-$17 57 56&   36.5$\pm$    3.2&   K2&   M2&          2011-05-29\\
 30 & 18 10 58.3& $-$17 41 24&   53.2$\pm$    0.8&   M0& $..$&          2011-08-19& 65 & 18 13 52.3& $-$17 57 54&   49.5$\pm$    1.0&   K5&   M7&          2011-05-29\\
 31 & 18 13 54.9& $-$17 53 49&   45.4$\pm$    6.8&   K5&   M6&          2011-06-29& 66 & 18 13 52.0& $-$17 57 52&   29.0$\pm$    8.1&   K1&   K5&          2011-05-29\\
 32 & 18 13 44.9& $-$17 57 12&   33.8$\pm$    2.8&   K2&   M1&          2011-08-19& 67 & 18 14 00.1& $-$17 56 31&   41.4$\pm$    2.0&   K3&   M4&          2011-05-29\\
 33 & 18 13 50.2& $-$17 54 26&   50.3$\pm$    1.2&   M0&   M7&          2011-05-31& 68 & 18 14 00.8& $-$17 56 51&   43.2$\pm$    1.0&   K4&   M5&          2011-05-29\\
 34 & 18 14 03.1& $-$17 58 52&   18.8$\pm$    2.2&   $<$K1&   $<$K1&          2011-08-20& 69 & 18 14 01.0& $-$17 56 50&   21.6$\pm$    3.9&   $<$K1&   K1&          2011-05-29\\
 35 & 18 14 03.2& $-$17 58 50&   44.0$\pm$    3.8&   K4&   M6&          2011-08-20& 70 & 18 14 09.4& $-$18 00 32&   37.6$\pm$    0.6&   K2&   M2&          2011-05-29\\
 36 & 18 14 03.4& $-$17 58 49&   42.0$\pm$    3.5&   K4&   M5&          2011-08-20& 71 & 18 10 52.5& $-$17 41 11&   51.2$\pm$    0.8&   M0&   M7&          2011-08-19\\
 37 & 18 14 02.7& $-$17 55 38&   49.0$\pm$    2.2&   K5&   M7&          2011-06-24& 72 & 18 13 52.7& $-$17 58 03&   39.7$\pm$    2.9&   K3&   M3&          2011-08-12\\
 38 & 18 13 49.9& $-$17 57 05&   43.7$\pm$    2.0&   K4&   M5&          2011-08-19& 73 & 18 13 52.9& $-$17 58 02&   27.1$\pm$    2.4&   $<$K1&   K3&          2011-08-12\\
 39 & 18 14 27.7& $-$17 57 05&   30.9$\pm$    1.6&   K1&   K5&          2011-06-24& 74 & 18 14 08.5& $-$18 00 19&   20.9$\pm$    7.3&   $<$K1&   K1&          2012-06-26\\
 40 & 18 14 06.3& $-$17 28 33&   18.9$\pm$    1.7&   $<$K1&   $<$K1&          2011-06-24& 75 & 18 14 08.8& $-$18 00 17&   21.2$\pm$    3.5&   $<$K1&   K1&          2012-06-26\\
 41 & 18 14 05.6& $-$17 28 50&   49.8$\pm$    1.1&   K5&   M7&          2011-06-24& 76 & 18 14 09.4& $-$18 00 48&   32.4$\pm$    1.8&   K1&   M0&          2012-08-13\\
 42 & 18 10 54.8& $-$17 39 56&   40.9$\pm$    2.2&   K3&   M4&          2011-06-07& 77 & 18 13 49.1& $-$17 56 15&   25.0$\pm$    1.3&   $<$K1&   K2&          2012-09-15\\
 43 & 18 10 55.1& $-$17 40 25&   31.8$\pm$    2.7&   K1&   K5&          2011-06-07& 78 & 18 13 55.4& $-$17 54 31&   24.8$\pm$    1.6&   $<$K1&   K2&          2012-09-02\\
 44 & 18 10 52.7& $-$17 40 19&   47.1$\pm$    1.1&   K5&   M7&          2011-06-07& 79 & 18 13 53.4& $-$17 55 19&   21.2$\pm$    2.6&   $<$K1&   K1&          2012-09-02\\
 45 & 18 10 52.7& $-$17 40 08&   44.1$\pm$    1.4&   K4&   M5&          2011-06-07& 80 & 18 13 52.2& $-$17 56 22&   42.4$\pm$    0.9&   K4&   M4&          2012-08-08\\
 46 & 18 10 50.5& $-$17 40 29&   23.7$\pm$    2.4&   $<$K1&   K2&          2011-06-07& 81 & 18 13 57.2& $-$17 58 08&   31.9$\pm$    2.6&   K1&   M0&          2012-08-11\\
 47 & 18 10 55.2& $-$17 41 20&   62.5$\pm$    9.0&   M2& $..$&          2011-05-20& 82 & 18 13 57.2& $-$17 58 06&   22.8$\pm$    6.6&   $<$K1&   K2&          2012-08-11\\
 48 & 18 10 56.6& $-$17 41 54&   44.4$\pm$    1.6&   K4&   M5&          2011-06-07& 83 & 18 13 38.1& $-$17 43 19&   46.5$\pm$    0.8&   K5&   M6&          2012-06-03\\
 49 & 18 14 01.8& $-$17 54 43&   46.7$\pm$    1.2&   K5&   M7&          2011-08-12& 84 & 18 13 20.8& $-$18 06 26&   44.6$\pm$    0.6&   K4&   M5&          2012-06-21\\
 50 & 18 14 03.9& $-$17 55 06&   52.2$\pm$    1.5&   M0& $..$&          2011-08-12& 85 & 18 13 13.5& $-$17 48 07&   54.6$\pm$    0.7&   M0& $..$&          2012-06-03\\
 51 & 18 13 54.7& $-$17 54 57&   56.1$\pm$    1.4&   M1& $..$&          2011-06-24& 86 & 18 14 44.5& $-$18 07 38&   50.2$\pm$    1.3&   K5&   M7&          2012-06-21\\
 52 & 18 13 53.4& $-$17 55 10&   14.2$\pm$    2.5&   $<$K1&   $<$K1&          2011-06-24& 87 & 18 13 48.2& $-$17 50 42&   57.5$\pm$    1.8&   M1& $..$&          2012-06-21\\
 53 & 18 13 52.9& $-$17 55 00&   53.4$\pm$    1.4&   M0& $..$&          2011-06-24& 88 & 18 13 54.8& $-$18 06 56&   49.9$\pm$    2.6&   M0&   M7&          2012-06-21\\
 54 & 18 13 52.5& $-$17 56 16&   60.0$\pm$    1.8&   M2& $..$&          2011-05-29& 89 & 18 14 07.8& $-$17 28 37&   55.4$\pm$    2.8&   M1& $..$&          2012-09-15\\
 55 & 18 13 51.9& $-$17 56 27&   44.9$\pm$    1.1&   K4&   M6&          2011-05-29& 90 & 18 14 02.9& $-$17 29 01&   40.0$\pm$    1.2&   K3&   M4&          2012-09-15\\
 56 & 18 13 55.5& $-$17 56 18&   43.4$\pm$    0.4&   K4&   M5&          2011-05-29& 91 & 18 14 05.5& $-$17 29 25&   30.9$\pm$    3.0&   K1&   K5&          2012-08-19\\
 57 & 18 13 54.6& $-$17 56 12&   57.1$\pm$    7.0&   M1& $..$&          2011-05-29& 92 & 18 14 10.1& $-$17 27 57&   29.4$\pm$    2.1&   K1&   K4&          2012-09-15\\
 58 & 18 13 54.1& $-$17 57 22&   40.8$\pm$    6.5&   K3&   M4&          2011-05-29& 93 & 18 13 47.4& $-$17 57 10&   41.3$\pm$    1.6&   K3&   M4&          2012-06-27\\
 \hline
\end{tabular}
\end{center}
\begin{list}{}{}
\item[{\bf Notes.}] ($^*$) Spectral types are estimated by using the relation between spectral types and
EW(CO)s of red giants (RGBs), as well as that between spectral types and
EW(CO)s of RSGs.
\end{list}
\end{table*}

\subsection{Extinction in \Ks-band and bolometric corrections}
\label{extinction}
For early-type stars, we assumed intrinsic colors, and effective temperatures, \Teff,
as tabulated per spectral-type in \citet{messineo11}, and based on the
works by \citet{bibby08}, \citet{crowther06bs}, \citet{johnson66}, 
\citet{koornneef83}, \citet{humphreys84}, \citet{lejeune01}, \citet{martins05},  
\citet{martins06}, and \citet{wegner94}. For the WN6,  we used the \Teff\ values 
and average infrared colors listed by  \citet{crowther07} and \citet{crowther06wd1}. 
For late-type stars, we adopted the intrinsic colors given by \citet{koornneef83}. 

Total extinction in \Ks-band was calculated by assuming these
intrinsic colors,  and by adopting the power-law curve with an index of $-1.9$
by \citet{messineo05}.
Estimates for the IR excess in three different colors  (E($J-H$), E($H-$\Ks), 
and E($J-$\Ks)) are provided in Table \ref{aktable}.
Since the \Ks-band  of mass-losing early-type stars (e.g., WR) may have significant 
excess due to free-free emission and dust \citep[][]{cohen75}, it is preferable to use  E$(J-H)$.
For late-type stars,  the E($J-$\Ks) is typically used.

\begin{figure}
\begin{center}
\resizebox{01.\hsize}{!}{\includegraphics[angle=0]{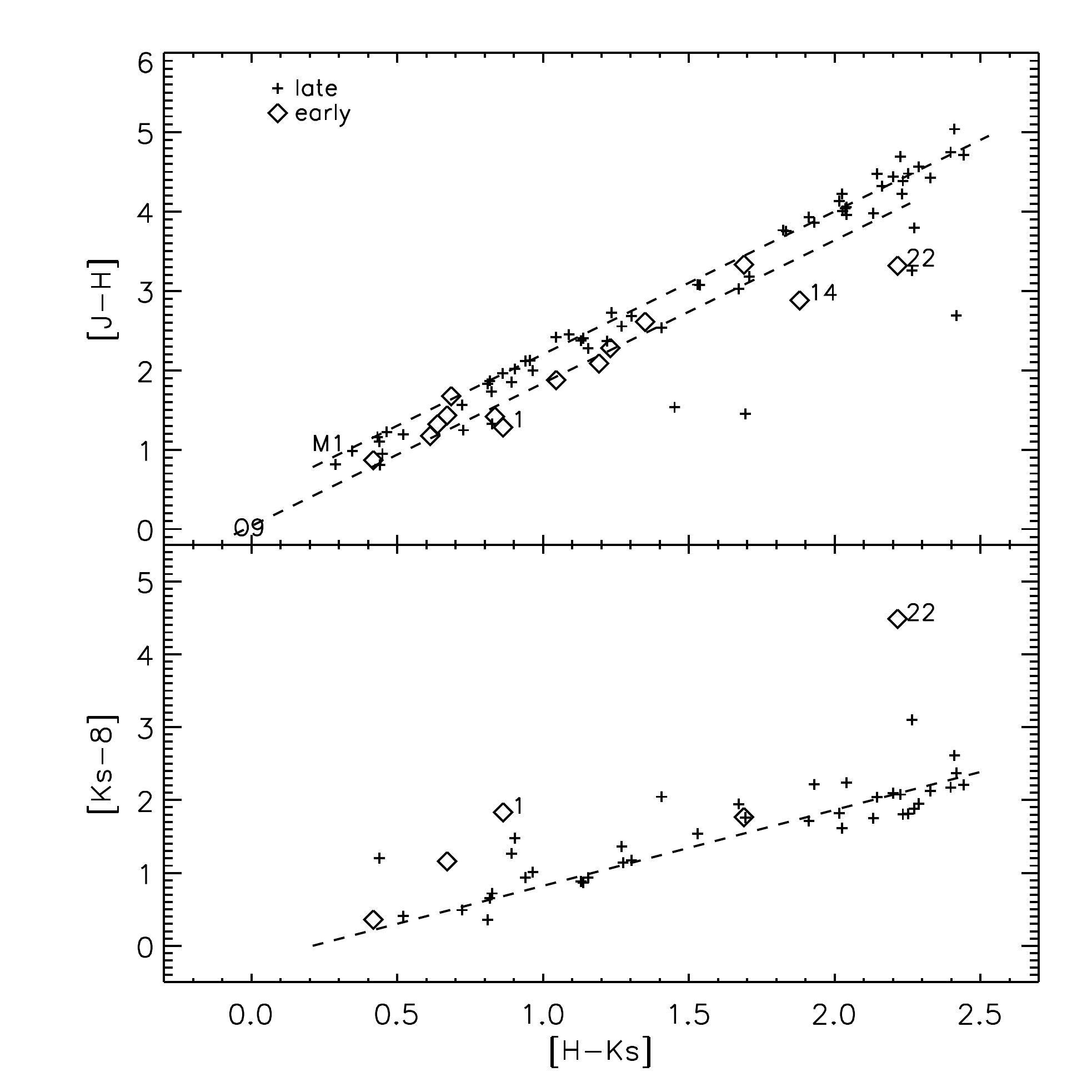}}
\caption{\label{colcol.eps} 
{\it Top panel:} $H-$\Ks\ versus $J-H$ colors of observed stars are shown.
The two dashed lines trace the locus of an M1 star (upper line) 
and O9 star (lower line) with increasing extinction \Aks\ from 0.0 mag to 3.5 mag.
{\it Bottom panel:} $H-$\Ks\ versus \Ks-8 colors. The dashed line trace the
locus of a star  ($H-$\Ks=0 mag, \Ks-8=0 mag) 
with increasing extinction \Aks\ from 0.0 mag to 3.5 mag. }
\end{center}
\end{figure}

Apparent bolometric magnitudes are calculated with dereddened \Ks\ magnitudes,
and bolometric corrections, \BCK, as listed in Tables 8, 9, and 10 by 
\citet[][ and references therein]{messineo11};
for the WN6, the adopted \BCK\ is taken  from  \citet{crowther06wd1}; for late-type stars
\BCK\ values per spectral type are available from the work of \citet{levesque05}.
As shown in the  $J-H$ versus $H-$\Ks\  diagram and in the \Ks$-8$ versus $H-$\Ks\ diagram 
of Fig. \ref{colcol.eps}, the bulk of the sources follow the direction expected for reddening by
interstellar dust. The two early-type stars \#14 (OBe) and \#22 (Oe) show significant 
infrared excess, possibly due to the presence of circumstellar material.


\begin{table*}
\begin{center}
\caption{\label{aktable} Color Properties of Newly Detected Early-type Stars.}
\begin{tabular}{rlrrrrrrrrr}
\hline
\hline
ID    &          Sp. Type &    $(J-K_S)_o$      & $(H-K_S)_o$      & A$_{\rm K_S}(JH)$     &  A$_{\rm K_S}(JK_S)$     &  A$_{\rm K_S}(HK_S)$    &    $Q1$   &  $Q2$ \\  
      &                   &     [mag]           & [mag             &   [mag]               &  [mag]                   &   [mag]                 &    [mag]  &  [mag] \\
\hline 
  1   &                 WN6   &   0.370     &  0.260     &  0.981     &  0.953     &  0.902     &                  $-$0.272$\pm$ 0.075     &                  $-$2.784$\pm$ 0.008    \\
  2   &                OBAF   &  $-$0.050     & $-$0.040     &  1.123     &  1.093     &  1.041     &                   0.149$\pm$ 0.043     &                               $..$    \\
  3   &                B0$-$5   &  $-$0.160     & $-$0.080     &  1.470     &  1.353     &  1.144     &                   0.442$\pm$ 0.137     &                               $..$    \\
  4   &                B0$-$5   &  $-$0.160     & $-$0.080     &  1.267     &  1.215     &  1.123     &                   0.224$\pm$ 0.077     &                  $-$1.017$\pm$ 0.086    \\
  5   &                OBAF   &  $-$0.050     & $-$0.040     &  0.337     &  0.375     &  0.444     &                  $-$0.071$\pm$ 0.045     &                               $..$    \\
  6   &                B0$-$5   &  $-$0.160     & $-$0.080     &  1.254     &  1.295     &  1.369     &                  $-$0.086$\pm$ 0.658     &                               $..$    \\
  7   &                O4$-$6   &  $-$0.210     & $-$0.100     &  0.819     &  0.803     &  0.775     &                   0.116$\pm$ 0.163     &                   0.318$\pm$ 0.010    \\
  8   &                O4$-$6   &  $-$0.210     & $-$0.100     &  2.883     &  2.809     &  2.676     &                   0.292$\pm$ 0.138     &                   0.270$\pm$ 0.143    \\
  9   &                OBAF   &  $-$0.050     & $-$0.040     &  2.722     &  2.682     &  2.612     &                   0.168$\pm$ 0.043     &                               $..$    \\
 10   &                B0$-$5   &  $-$0.160     & $-$0.080     &  1.978     &  1.972     &  1.961     &                   0.065$\pm$ 0.051     &                               $..$    \\
 11   &                O4$-$6   &  $-$0.210     & $-$0.100     &  2.280     &  2.240     &  2.169     &                   0.182$\pm$ 0.107     &                               $..$    \\
 12   &                OBAF   &  $-$0.050     & $-$0.040     &  0.313     &  0.243     &  0.117     &                   0.296$\pm$ 0.054     &                               $..$    \\
 13   &                B0$-$5   &  $-$0.130     & $-$0.030     &  1.067     &  1.029     &  0.962     &                   0.071$\pm$ 0.059     &                               $..$    \\
 14   &                 OBe   &  $-$0.120     & $-$0.060     &  2.465     &  2.622     &  2.902     &                  $-$0.502$\pm$ 0.061     &                               $..$    \\
 15   &                OBAF   &  $-$0.060     & $-$0.010     &  0.658     &  0.560     &  0.386     &                   0.289$\pm$ 0.054     &                               $..$    \\
 16   &                OBAF   &  $-$0.060     & $-$0.010     &  0.580     &  0.518     &  0.408     &                   0.169$\pm$ 0.063     &                               $..$    \\
 17   &                B0$-$5   &  $-$0.130     & $-$0.030     &  1.832     &  1.831     &  1.828     &                  $-$0.059$\pm$ 0.045     &                               $..$    \\
 18   &                B0$-$5   &  $-$0.130     & $-$0.030     &  1.657     &  1.639     &  1.608     &                  $-$0.003$\pm$ 0.051     &                               $..$    \\
 19   &                OBAF   &  $-$0.050     & $-$0.040     &  0.607     &  0.511     &  0.340     &                   0.378$\pm$ 0.051     &                               $..$    \\
 20   &                OBAF   &  $-$0.050     & $-$0.040     &  0.529     &  0.491     &  0.423     &                   0.184$\pm$ 0.048     &                               $..$    \\
 21   &                OBAF   &  $-$0.050     & $-$0.040     &  0.602     &  0.591     &  0.570     &                   0.095$\pm$ 0.046     &                               $..$    \\
 22   &                  Oe   &  $-$0.210     & $-$0.100     &  2.872     &  3.085     &  3.465     &                  $-$0.670$\pm$ 0.129     &                  $-$6.532$\pm$ 0.344    \\
 23   &                O6$-$7   &  $-$0.210     & $-$0.100     &  1.199     &  1.165     &  1.104     &                   0.173$\pm$ 0.083     &                               $..$    \\

\hline
\end{tabular}
\begin{list}{}{}
\item[{\bf Notes.}] The $Q1$ and $Q2$ parameters are defined as in \citet{messineo12}.
\end{list}
\end{center}
\end{table*}

\section{Stellar parameters of W33 members}

In the following, we discuss  the properties of the detected massive 
early-type stars and their association with W33.
Recently, \citet{immer13}  measured parallactic distances of several water masers
in the direction of  W33. 
 The centroid LSR velocities of 3 out of 4 H$_2$O maser sites are 
between 29 and 37 km~s$^{-1}$,  while that of the remaining,  W33B, is 59.3 km~s$^{-1}$. 
Nevertheless, the trigonometric parallaxes yield similar values for their distances.

Their average distance  is $2.64$ kpc with a standard deviation $\sigma =0.25$ kpc.
Following Immer et al., for the entire W33 complex we adopt the parallactic distance 
to the maser W33B of $2.4^{+0.17}_{-0.15}$ kpc  ($DM=11.90\pm0.16$ mag).
Derived  absolute \Ks, \Mk, and bolometric magnitudes, 
\Mbol\ are listed in  Tables \ref{aktable} and \ref{table:mbol}.

\subsection{O-type stars}
We  detected three luminous  O stars within  W33 as part of our survey 
- \#7, \#8 (O4-6), and \#23 (O6-7) - which
are  amongst the brightest stars appearing in the \Ks\ versus $H-$\Ks\
diagrams of Fig.\ \ref{CMDs}. 

Stars \#7 and \#8 are located in the Mercer1 region, which 
coincides with   the \HH\ region G12.745$-$00.153 
\citep{lockman89,white05} -- in SIMBAD this  object is named [L89b]12.745$-$00.153.
The two O-type stars  have a angular separation of 1\farcm3; 
the first has an \Aks=  $0.82\pm0.05$ mag, the latter has an \Aks\ more than 3 times larger.
 The location of star \#7  near the peak  of the 24 \um\ emission of the \HH\ region
\citep[][]{carey09}  provides evidence for its association with W33;
star \#8 is located in the dustier surrounding 8 \um\ shell (visible in GLIMPSE).
Similar variations of \Aks\ are reported within other mid-IR bubbles   
\citep[e.g.,][]{bik10}; therefore, we  attribute the difference of 
interstellar extinction between the two stars to strong dust variations of the 
same \HH\ in W33. Star \#23 is located in the cl1 region, and has an \Aks\ value 
of $ 1.20\pm0.03$ mag (similar to that of star \#7). 

By assuming a common distance of 2.4 kpc, bolometric corrections as listed in Table 5,
and a solar bolometric constant of $-4.74$ mag,
we derived log(L/\Lsun)=$ 5.51^{+0.09}_{-0.09}$, $ 5.61^{+0.09}_{-0.09}$, 
and $ 5.56^{+0.07}_{-0.07}$, 
and \Mk=$ -4.63\pm0.17$ mag, $ -4.89\pm0.18$ mag, and $ -5.00\pm0.16$ mag, 
for stars \#7, \#8, and \#23, respectively.
The similarity of \Mbol\ (and \Mk) values suggests that these three stars 
have similar ages and masses.  Supporting our assertions  in the preceding section 
regarding their likely evolved nature, comparison with 
\Mk\ values of mid O-type stars  \citep{martins06},
indicates that all three are consistent with luminosity classes III-I. 

By using the latest stellar models by the Geneva group  with solar metallicity
and rotation \citep{ekstrom12}, we derive stellar masses from 30 \Msun\ to 40 \Msun, 
and an age below 6 Myr, at which point all 40 \Msun\ stars would have been lost to SNe; 
the presence of spectroscopically  O4-6 supergiants  (\Mk=$ -5.03$ mag with 
$ \sigma=0.47$ mag
\footnote{ The average \Mk\ is calculated with the magnitudes from \citet{figer02}, 
a distance of 8.4 kpc, and the exctinction law by \citet{messineo05}.}) within the Arches 
\citep[2-4 Myr,][]{martins08}  and of O4-6 dwarfs 
(\Mk=$ -4.15$ mag with $ \sigma = 0.43$ mag 
\footnote{ The average \Mk\ is calculated with the magnitudes 
and  distance provided by \citet{davies12} and the exctinction law by \citet{messineo05}.}) in Danks 1 
\citep[$\sim1.5^{+1.5}_{-0.5}$ Myr,][]{davies12} suggests 
an age of 2-4 Myr.

\begin{figure*}
\begin{center}
\resizebox{0.99\hsize}{!}{\includegraphics[angle=0]{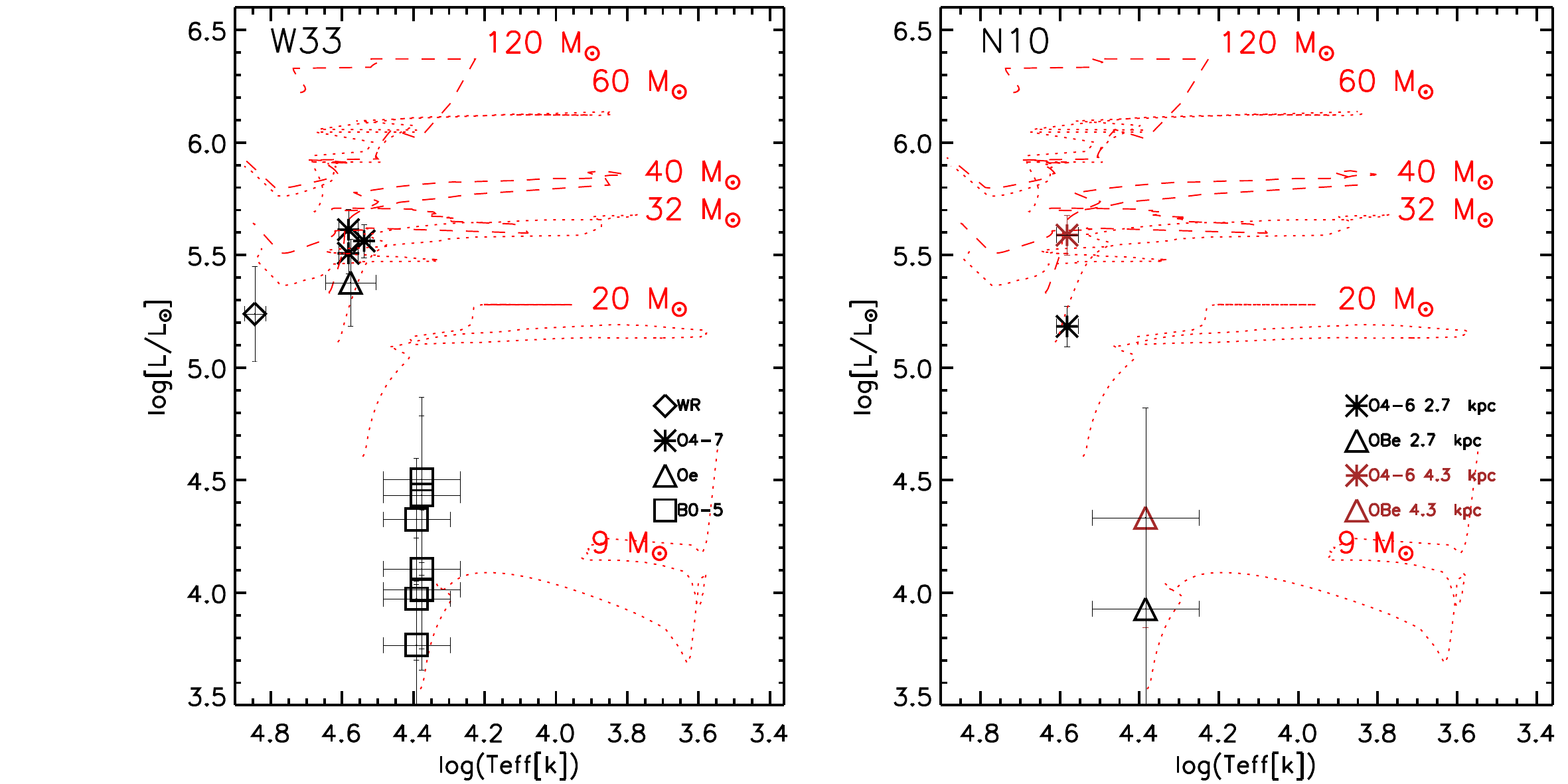}}
\caption{\label{luminosity}
Luminosity versus \Teff\ diagram of massive O-type stars (O4-6 and O6-7), B0-5 stars,
and of the WN6b star detected in the direction of W33 ({\it left}) and of N10 ({\it right}).
For W33, we assume a distance of 2.4 kpc; for N10, the same two stars are plotted for a distance of
2.4 kpc (black points) and for a distance of 4.3 kpc (lighter points).
Stellar tracks for stars of 9, 20, 32, 40, 60, and 120 \Msun, 
with solar metallicity and including rotation, 
are shown with dotted and dashed curves  \citep{ekstrom12}.}
\end{center}
\end{figure*}

At the position of Oe star \#22, the SINFONI cube  shows  several H$_2$ lines, 
suggesting that star \#22 (\Ks=10.448 mag) is still embedded. 
We derived \Aks=$ 2.87\pm0.07$ mag from the $J-H$ color excess;   by assuming 
a distance of 2.4 kpc, we estimate \Mk=$ -4.33\pm0.18$ mag.
For confirming its luminosity class, further spectroscopy in $J$ and $H$-band is
required.

\subsection{A new WN6}

Found  in the south-west periphery of W33 (as shown in Fig. 7), the WN6b star
\#1 has an \Aks=1.0 mag, when assuming the average intrinsic near-infrared
colors  for late WN  stars by \citet{crowther06wd1}. With a distance of 2.4 kpc  
we measured $M_K= -4.86\pm0.16$ mag, which fits well with the average 
$M_K=-5.13$ mag ($\sigma=0.07$ mag) of two other  WN6b stars analyzed by \citet{crowther06wd1}.
By using  \BCK=$-3.5\pm0.5$ mag we obtain log(L/\Lsun)=$5.24^{+0.21}_{-0.21}$, 
and a mass of about  $27\pm2.5$ \Msun.  The detection of a late WN  implies the existence
of highly luminous  progenitor supergiants \citep{georgy12}, 
such as those we detected in Mercer1.

The compilations of Galactic WR stars by \citet{vanderhucht01}, \citet{mauerhan11}, 
\citet{shara12}, and \citet{faherty14}, identify   43  WN6 stars from a total of 
443 WRs of all flavours. Among the 12 WN6 stars listed in the latter three works 
40\% have broad features, suggesting that only 4\% of known WRs
have a similar classification. Rapid rotation and the presence of a magnetic field 
have been suggested to  explain the broadening  of their spectral lines and the flattening 
of the line peaks \citep{shenar14}.

\subsection{Spectral types B0-5} 
Six B0-5 stars were detected. Stars \#3, \#4, \#6, \#10, and \#13 
are located in the  Mercer1 region, with \Aks\ values of $ 1.47\pm0.04$ mag,  
$ 1.27\pm0.03$ mag, 
$ 1.25\pm0.21$ mag, $ 1.98\pm0.02$ mag, and $ 1.07\pm0.02$ mag, respectively.
For a distance of 2.4 kpc, their \Mk\ values range from $-2.07$ mag to $-3.69$ mag,
and suggest a mix of  dwarfs and giants 
{ \citep{martins06,humphreys84,wegner94, lejeune01} } 
with initial masses from 9 to  15 \Msun\ \citep{ekstrom12}.

The B0-5 stars \#18 and \#17 are located in the cl2 cluster. Star \#18 is the brightest 
star in the small cluster. For a distance of 2.4 kpc, it has \Aks=$ 1.66\pm0.02$ mag 
and \Mk $= -2.96\pm0.16$ mag,
which suggests a luminosity class V/III and an initial mass of $12\pm3$ \Msun\ \citep{ekstrom12}.
Star \#17, with \Aks =$ 1.83\pm0.01$ mag and \Mk=$ -1.56\pm0.16$ mag, has a likely initial mass 
of $10\pm2$ \Msun.

We note that the nine stars generically classified as `early' (i.e., those assigned 
spectral type OBAF in Table 1) exhibit such an uncertainty in  temperatures and hence 
intrinsic colours that we cannot derive meaningful physically properties for them at this time.

\subsection{Late-type Stars}
For a distance  of 2.4 kpc, the detected late-type stars from Table \ref{table:obslate} 
remain fainter than \Mbol=$-$5.26 mag  (log(L/\Lsun)=4.0). They  all have magnitudes consistent with those
of giant stars.

\section{The global structure and star formation history of  W33}
\label{w33main}

\begin{figure}
\begin{center}
\resizebox{0.99\hsize}{!}{\includegraphics[angle=0]{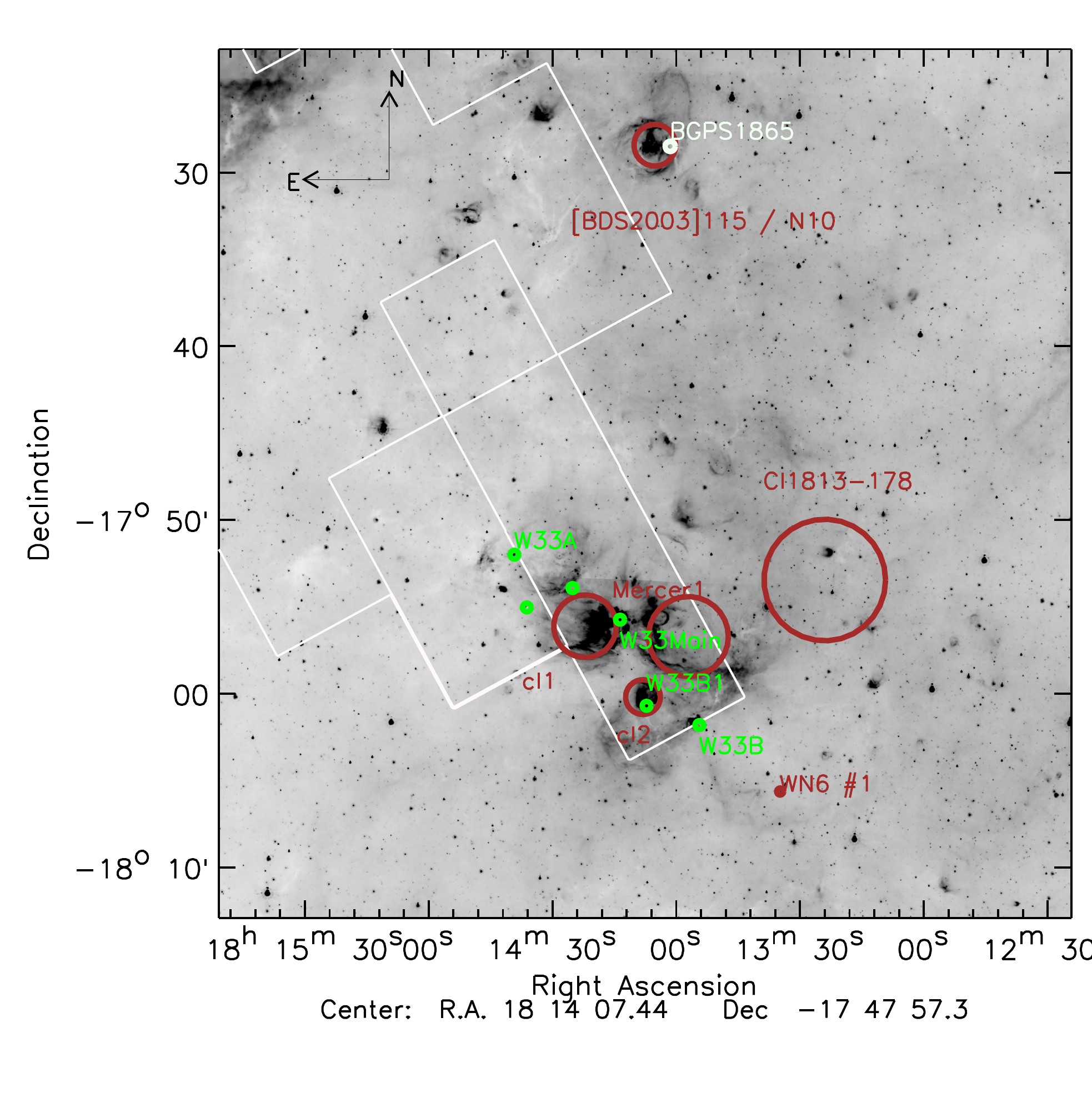}}
\caption{\label{fullmap}
 8 \um\ emission from GLIMPSE \citep{spitzer09} of W33.
The Mercer1, cl1, cl2 and Cl1813$-$178 regions are marked with circles, 
as well as the isolated WN6 star \#1. 
The green dots indicate the location of the molecular clumps  in W33
detected by \citet{immer14},  and in blue that in N10 studied by \citet{ma13}.
The white contours show the location of the peak of $^{12}$CO emission in the range
from 30  \kms\ to 60 \kms\ from the Galactic survey of \citet{dame01}.
The image has a linear size of 34.5 pc $\times$ 35.09 pc for a distance of 2.4 kpc.
}
\end{center}
\end{figure}

Complementing Fig. \ref{strange}, in which we show the location of (candidate) massive stars in the 
putative individual clusters, Fig. \ref{fullmap} delineates the nominal locations of these regions
on a map of the 8 \um\  emission. 
Massive stars are found throughout W33, with  the richest region being 
the Mercer 1\footnote{Synonymous with the H II  G12.745$-$00.153.} aggregate to the west of W33 Main.
The presence of two O4-6 (super-)giants \#7 and \#8  suggests a burst of star formation 
occurred $\sim2-4$ Myr ago. A further five  early to mid B-type stars  
(\#3, \#4, \#6, \#10, and \#13)  have  \Aks\ consistent with  those of the O-type stars
and bolometric magnitudes typical of dwarfs and/or giants. The remaining three stars  
(\#2, \#5, and \#9) are of generic early (OBAF) spectral-type.

Immediately to the west of Mercer 1 and the embedded protocluster forming within the radio source W33 Main, 
we find the O4-6 (super-)giant \#23, which demonstrates
similar physical properties to stars \#7 and \#8.
In GLIMPSE images,  star \#23 is surrounded by a yellow curved filament,
similar to the mid-infrared bow shocks identified by \citet{povich08}.
The likely young massive Oe star \#22  is located in front of the apex of the bowshock 
between \#23 and the radio source W33 Main, consistent with it’s elevated extinction (\Aks=$ 2.87\pm0.07$ mag). 
The lack of further massive stars in this region leads us to conclude that the cl1 region
does not delineate a {\em bona fide} cluster.

\begin{figure*}
\begin{center}
\resizebox{0.49\hsize}{!}{\includegraphics[angle=0]{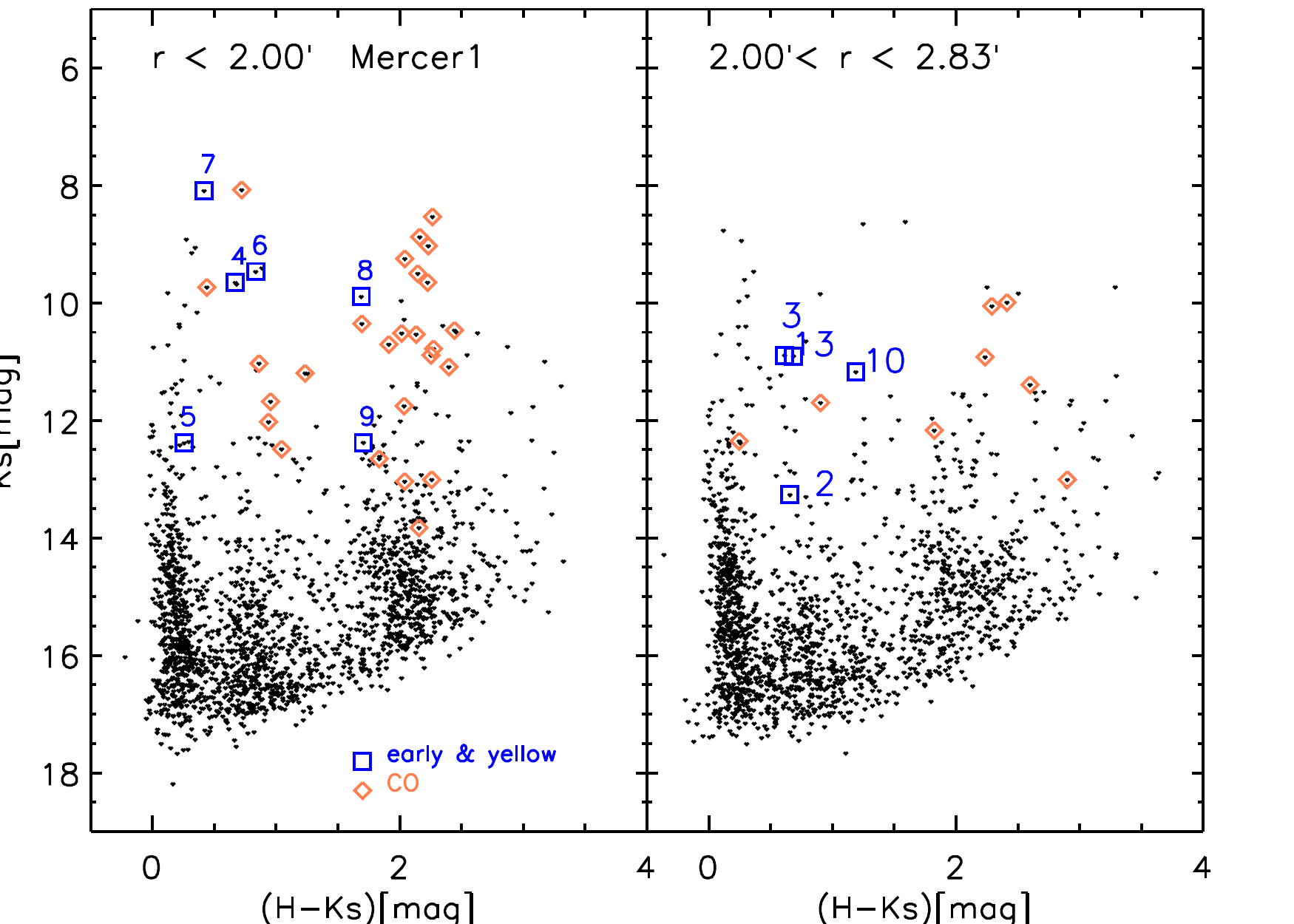}}
\resizebox{0.49\hsize}{!}{\includegraphics[angle=0]{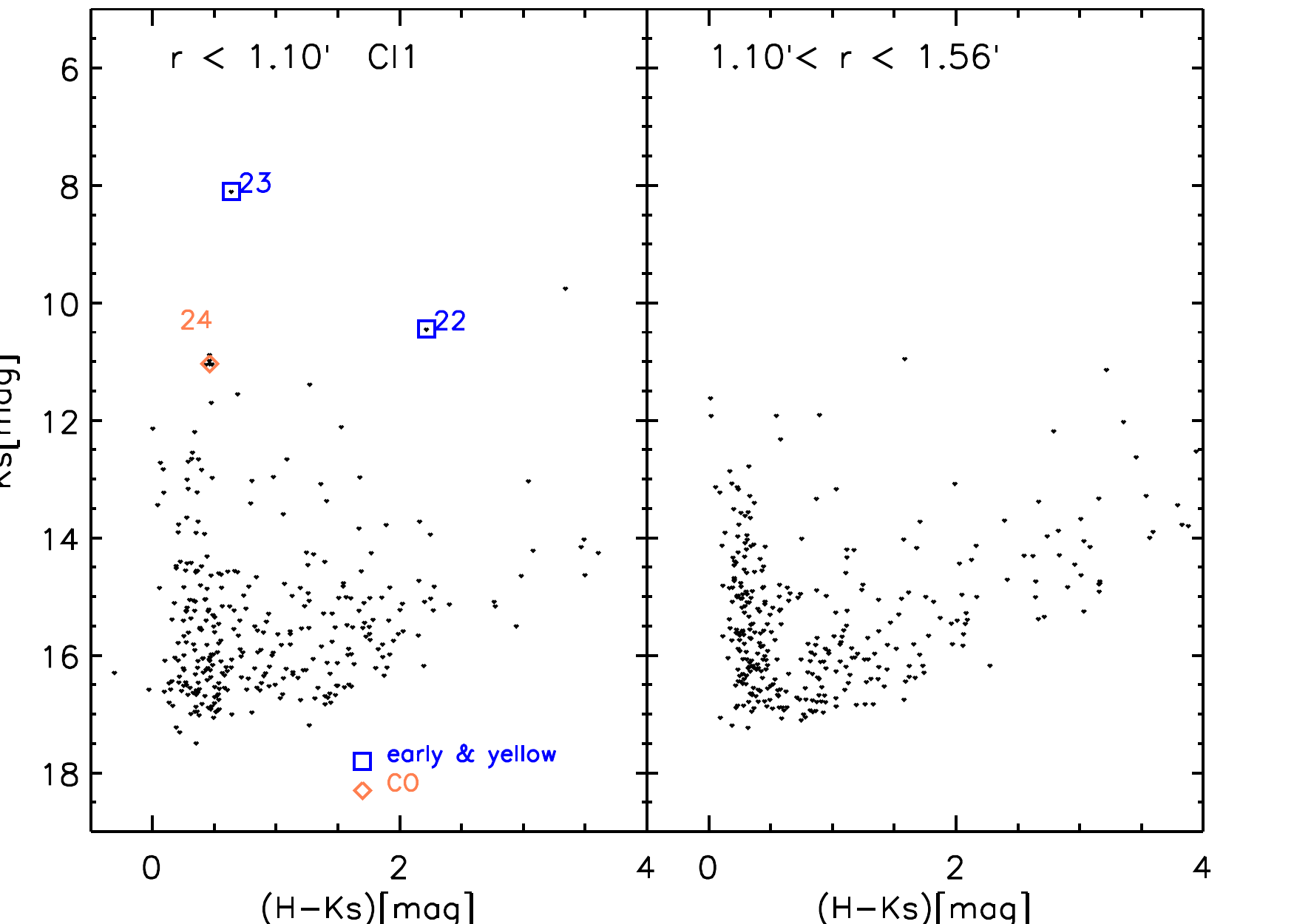}}
\end{center}
\begin{center}
\resizebox{0.49\hsize}{!}{\includegraphics[angle=0]{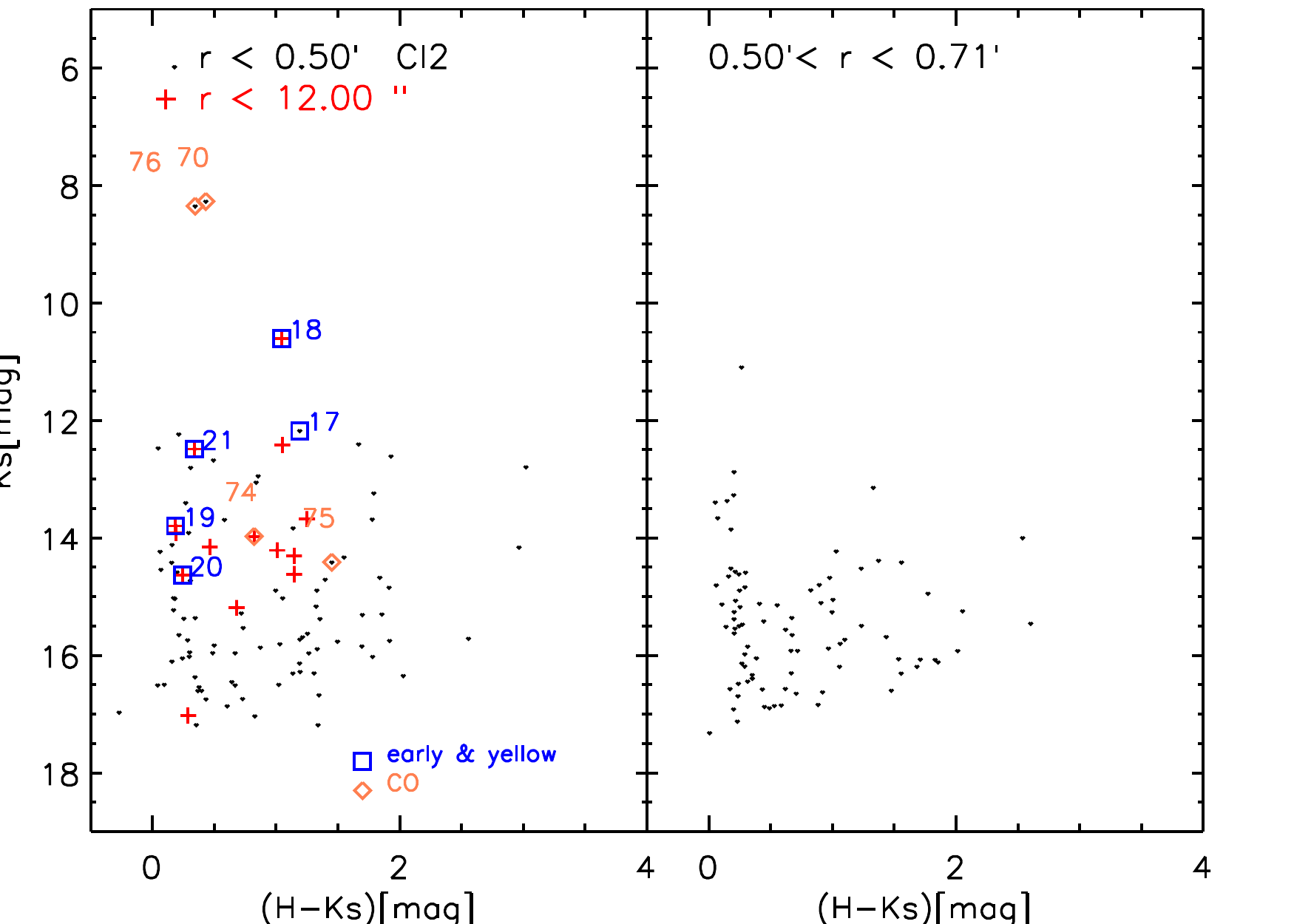}}
\resizebox{0.49\hsize}{!}{\includegraphics[angle=0]{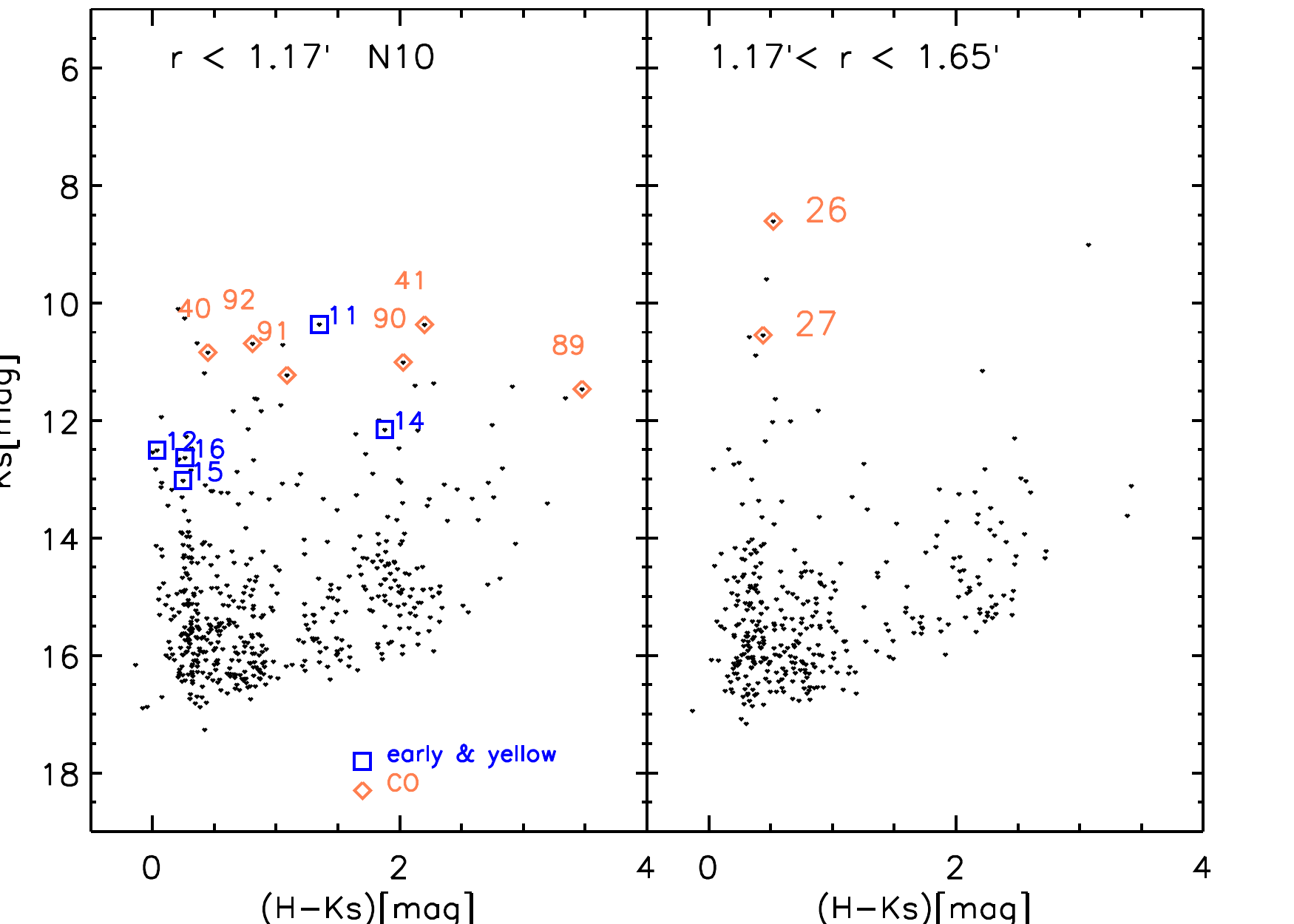}}
\end{center}
\caption{ \label{CMDs}  2MASS-UKIDSS $H-$\Ks\ versus \Ks\ diagram  
of the Mercer1 ({\it top left}), cl1 ({\it top right}),
cl2 ({\it bottom left }), and N10 ({\it bottom right })
 regions. Spectroscopic detected  early-yellow stars are 
indicated with squares, late-type stars with diamonds.
Identification numbers are taken from Tables \ref{table.obspectra}
and  \ref{table:obslate}. A comparison field of equal area is shown in the right panel;
field datapoints are taken from an annular region.
In the cl2 region,  stars concentrated in the central 12\arcsec\ are marked with crosses.} 
\end{figure*}

The massive protostar W33A \citep{davies10} is  located in the north-east of these regions,
while the stellar aggregate cl2 is located to the south.  A sequence of reddened stars 
is detected within the compact nebula of cl2 (Fig. \ref{strange} and Fig.\ \ref{CMDs}); 
these are found  at $J-$\Ks$\approx 1.6$ 
mag in the color magnitude diagram shown in Fig.\ \ref{CMDs}. The two brightest stars, 
\#17 and \#18, are spectroscopic B0-5 types, while the remaining three, \#19, \#20 and \#21, 
are classified as spectral type OBAF. Radio continuum emission is found in the direction  
of the cl2 cluster, centered on star \#18; we estimate a flux density  of 0.57 Jy at 20 cm, 
and 0.64 Jy at 90 cm by using MAGPIS data and an aperture of 35\arcsec. 
Under the assumption of optically thin thermal emission with an electron temperature, 
T$_{\rm e}$=10,000 K, this implies  a  Lyman continuum photon flux, N$_{\rm lyc}$, 
of 10$^{47.2}$ s$^{-1}$ \citep[e.g.,][]{martinh03,rubin68,storey95}.  For comparison, 
a O9.5 V emits a number of  N$_{\rm lyc}$ of $10^{47.9}$ s$^{-1}$ and a O9.5 III of
$10^{48.4}$ s$^{-1}$ \citep[from the more recent work by][]{martins05}.
By comparing the results from  \citet{martins05}, \citet{vacca96}, and 
\citet{panagia73}, after having corrected for relative average shifts, 
we estimate N$_{\rm lyc}$=10$^{47.2}$ s$^{-1}$, 10$^{48.0}$ s$^{-1}$,
10$^{44.4}$ s$^{-1}$, and  10$^{45.1}$ s$^{-1}$
for a B0 V, a B0 III, a B2 V, and a B2 III star, respectively. We
find typical uncertainties of 0.2 dex for every  N$_{\rm lyc}$ value.
Therefore,  stars \#17 and \#18 may already account for the requisite ionising flux.
While early B-type stars with initial masses of 9-12 \Msun\ are present 
in stellar populations with ages ranging up to 30 Myr \citep{ekstrom12}, 
the nebular emission associated with cl2  suggests  a much younger age, likely of only a few Myr. 

Finally, further to the south-west we find the broad lined WN6 star \#1. 
We unsuccessfully searched the W33 area for other possible bright stars (\Ks $<10$ mag) 
with properties of free-free emitters or candidate red supergiants by using the
infrared photometric criteria of \citet{messineo12}. Thus,  it is unlikely that 
there are any further young massive stellar aggregates associated with the complex. 
Consideration of these findings emphasizes that the massive stellar population is  
distributed across the confines of the W33 and appears not to be concentrated in 
rich young clusters similar to e.g., Danks 1 and Danks 2 within the G305 star forming complex \citep{davies12}. 

This behavior appears to mirror the current location of cold molecular material, within which 
future generations of 
stars may form. \citet{immer14} report the detection of six molecular clumps along the east side of W33, 
with masses of $0.2 - 4.0\times 10^{3}$ \Msun\, coincident with  the
 peak of the CO intensity map  (see Figure \ref{fullmap}). By contrast the evolved H II region  
 G12.745-00.153 (Mercer1)
resides on the west side of W33, where the molecular matter
has already been swept out, a configuration that is at least  suggestive of sequential star formation.
Similarly the dense clump W33B1 \citep{immer14} is located $\sim 35$\arcsec\ South-West 
of the cl2 cluster on the periphery of the apparent wind blown nebula associated with the latter; 
it is conceivable that its influence is contributing  
to this new protostellar core by heating and compressing  it \citep{immer14}.

\begin{table*}
\begin{center}
\caption{\label{table:mbol} Physical properties of newly detected massive stars. }
\begin{tabular}{rllrrrrrrlrrr}
\hline
\hline
ID$^b$   &   Sp. &  Class$^c$ & \Teff & $K_{\mathrm S_o}$  & \Aks &  $BC_{\mathrm K_S}$ &    $M_{ K_{\mathrm S}}$(der.) & $ {\mathrm log}(L/\Lsun)$ & Region \\
         &       &        & [K] &[mag]       &[mag] &  [mag]              &    [mag]                      &[mag] &\\
\hline
  1 &       WN6 &     I &  70000$\pm$  5000 & 7.043 $\pm$ 0.033 &  0.981 $\pm$ 0.025 & $-$3.50 $\pm$  0.50  & $-$4.86 $\pm$  0.16  &   $ 5.24 ^{+ 0.21}_{- 0.21  }$    &                 South of W33 \\
  3 &      B0$-$5 &   III &  24000$\pm$  6700 & 9.437 $\pm$ 0.059 &  1.470 $\pm$ 0.040 & $-$2.83 $\pm$  0.87  & $-$2.46 $\pm$  0.17  &   $ 4.01 ^{+ 0.35}_{- 0.35  }$    &              Mercer1, W33 \\
  4 &      B0$-$5 &   III &  24000$\pm$  6700 & 8.386 $\pm$ 0.035 &  1.267 $\pm$ 0.025 & $-$2.83 $\pm$  0.87  & $-$3.52 $\pm$  0.16  &   $ 4.43 ^{+ 0.35}_{- 0.35  }$    &              Mercer1, W33 \\
  6 &      B0$-$5 &   III &  24000$\pm$  6700 & 8.212 $\pm$ 0.245 &  1.254 $\pm$ 0.208 & $-$2.83 $\pm$  0.87  & $-$3.69 $\pm$  0.29  &   $ 4.50 ^{+ 0.37}_{- 0.37  }$    &              Mercer1, W33 \\
  7 &      O4$-$6 &     I &  38000$\pm$  2500 & 7.271 $\pm$ 0.057 &  0.819 $\pm$ 0.051 & $-$4.40 $\pm$  0.15  & $-$4.63 $\pm$  0.17  &   $ 5.51 ^{+ 0.09}_{- 0.09  }$    &              Mercer1, W33 \\
  8 &      O4$-$6 &     I &  38000$\pm$  2500 & 7.009 $\pm$ 0.067 &  2.883 $\pm$ 0.058 & $-$4.40 $\pm$  0.15  & $-$4.89 $\pm$  0.17  &   $ 5.61 ^{+ 0.09}_{- 0.09  }$    &              Mercer1, W33 \\
 10 &      B0$-$5 &   III &  24000$\pm$  6700 & 9.207 $\pm$ 0.020 &  1.978 $\pm$ 0.015 & $-$2.83 $\pm$  0.87  & $-$2.69 $\pm$  0.16  &   $ 4.11 ^{+ 0.35}_{- 0.35  }$    &              Mercer1, W33 \\
 13 &      B0$-$5 &     V &  25000$\pm$  5900 & 9.827 $\pm$ 0.026 &  1.067 $\pm$ 0.016 & $-$3.12 $\pm$  0.66  & $-$2.07 $\pm$  0.16  &   $ 3.97 ^{+ 0.27}_{- 0.27  }$    &              Mercer1, W33 \\
 17 &      B0$-$5 &     V &  25000$\pm$  5900 &10.343 $\pm$ 0.021 &  1.832 $\pm$ 0.013 & $-$3.12 $\pm$  0.66  & $-$1.56 $\pm$  0.16  &   $ 3.77 ^{+ 0.27}_{- 0.27  }$    &                  cl2, W33 \\
 18 &      B0$-$5 &     V &  25000$\pm$  5900 & 8.946 $\pm$ 0.023 &  1.657 $\pm$ 0.015 & $-$3.12 $\pm$  0.66  & $-$2.96 $\pm$  0.16  &   $ 4.33 ^{+ 0.27}_{- 0.27  }$    &                  cl2, W33 \\
 22 &        Oe &     $..$&  $..$             & 7.576 $\pm$ 0.075 &  2.872 $\pm$ 0.071 & $..$                 & $-$4.33 $\pm$  0.18  &   $..$                            &                  cl1, W33 \\
 23 &      O6$-$7 &     I &  35000$\pm$  1200 & 6.904 $\pm$ 0.037 &  1.199 $\pm$ 0.027 & $-$4.17 $\pm$  0.08  & $-$5.00 $\pm$  0.16  &   $ 5.56 ^{+ 0.07}_{- 0.07  }$    &                  cl1, W33 \\
 \hline
11 &      O4-6 &        I &  38000$\pm$  2500 & 8.083 $\pm$ 0.039 &  2.280 $\pm$ 0.032 & -4.40 $\pm$  0.15  & -5.08 $\pm$  0.10  &   $ 5.69 ^{+ 0.07}_{- 0.07  }$    &             BDS2003-115 \\
14 &       OBe &   $..$   &  $..$             & 9.691 $\pm$ 0.028 &  2.465 $\pm$ 0.026 & $..$                 & $-$3.47 $\pm$  0.10  &   $..$                            &          BDS2003-115 \\

\hline
\end{tabular}
\begin{list}{}{}
\item[{\bf Notes.}] A  distance of $2.4^{+0.17}_{-0.15}$  kpc 
(DM = $+11.90\pm0.16$ mag) is used for W33 \citep{immer13},
and of  $4.29^{0.17}_{-0.19}$ kpc (DM = $13.16\pm 0.09$ mag)  for N10.
($^b$) OBAF detections are not included in this table. 
 ($^c$) Classes are photometrically estimated 
using   \Mk\ values from  \citet{martins06}, \citet{bibby08}, \citet{humphreys84}, 
and \citet{lejeune01}. 
\end{list}
\end{center}
\end{table*}

\section{BDS2003-115 and Bubble N10}
\label{N10}

The cluster candidate BDS2003-115 is located North of W33 \citep{bica03, messineo11}.  
It appears as a group of  bright near-infrared stars in the core of a mid-infrared bubble 
\citep[bubble N10, and candidate SNR G13.1875+0.0389,][]{helfand06,watson08,churchwell06}\footnote{The 
center of the bubble  is dominated by radio continuum emission and 
8 \um\ and 24 \um\ emission; the latter likely derived from warm dust not yet destroyed by stellar feedback.  
With MAGPIS data \citep{helfand06}, we estimated 
an integrated flux density of  5.3 Jy at 20 cm, and of 7.5 Jy at 90 cm, 
over identical areas of 2\arcmin\ radii. The resulting spectral index $\alpha\sim-0.2$. 
Such a value is marginally consistent with that
expected from the optically thin free-free emission ($\alpha\sim-0.1$) 
expected for an HII region, although  one cannot exclude an additional 
source of non-thermal emission, 
originating from either the shocked stellar winds from massive OB stars 
or a supernova explosion \citep{leitherer97,williams96, sidorin14}.}.
Radio line observations of the $^{13}$CO $J=3-2$ transition yielded a radial 
velocity \Vlsr=  $50.2\pm4.1$ \kms\  \citep{beaumont10}. 
 We calculated the distance to N10 using the Galactic rotation curve 
parameters determined by \citet{reid09} from fitting trigonometric parallaxes 
of star forming regions, i.e., $R_\odot = 8.4\pm0.6$ kpc and $\Theta_0=254\pm16$ \kms.  
We find a distance of 4.29$^{+0.17}_{-0.19}$ kpc.
This could correspond to the high-velocity component seen in direction of W33; 
the  low-velocity component of W33  (35 \kms) is also detected along the 
line-of-sight of N10 \citep{dame01}.
Bubble  N10 is among the 28\% of bubbles currently interacting with a molecular condensation \citep{deharveng10};
this  clump (hereafter, BGPS1865) is locate on the western edge of the bubble and  has a mass ($>1600$ \Msun) 
\citep[][and references therein]{ma13}.

We detected one  O4-6 star (star \#11, \Aks=$ 2.28\pm0.03$ mag) 
in the center of the bubble N10, as well as 3  early type-stars 
(spectral-type OBAF; \#12, \#15, and \#16) and  the  embedded massive star \#14.
For a distance of 2.4 kpc, we obtain \Mk=$-4.07\pm0.1$ mag for \#11
\citep[which is a typical value for dwarfs of $ 29\pm3$ \Msun,][]{martins06,ekstrom12};
for a kinematic distance of  4.29 kpc this is revised upwards to \Mk=$-5.08\pm0.1$ mag
\citep[(super-)giant of $ 36\pm4$ \Msun,][]{martins06,ekstrom12}. 
There is a hint for Si IV  in emission at 2.428 \um\ in the spectrum of star \#11.
Given the uncertainty in reddening  and temperature of the remaining 4 stars  
we are unable to determine their luminosities.
 
\section{The relationship between Cl1813$-$178 and W33}
\label{cl1813}

The young massive  cluster, Cl 1813$-$178 is projected onto the north-west periphery of W33, 
and is coincident with  SNR  G12.72$-$0.00 \citep[e.g.,][]{helfand06,brogan06,messineo08,messineo11}.
Messineo et al. also suggested a possible association of this cluster,
with G12.83$-$0.02, and the pulsar and TeV Gamma-Ray 
Source PSR J1813$-$1749/HESS J1813$-$178 \citep[][]{brogan05}.

The cluster contains  six spectroscopically detected late O-type stars, twelve early 
B-type stars, two WN7 stars, and three transitional objects \citep[the O6O7If star \#5, 
the O8O9If \#16, and the cLBV  \#15 in ][]{messineo11}.
\citet{messineo11} estimated a spectrophotometric distance of $3.7\pm1.7$ kpc, 
consistent within errors with a stellar kinematic distance of $ 4.8\pm0.2$ kpc, 
which is based on the radial velocity of the RSG member \citep[\Vlsr = $ 62\pm4$ \kms,][]{messineo08}
 and on the Galactic rotation parameters presented by \citet{reid09}. 

Given our current understanding of stellar evolution, it  appears difficult  to reconcile 
the distance of Cl 1813$-$178 with the new parallactic distance to  the W33 complex ($\sim2.4$ kpc). 
On the basis of the spectral features, luminosity classes can be  inferred 
only for the transitional objects, the two WRs, and 
two other B0-B3 stars with He I at 2.058 \um\ in emission.
While the magnitudes of O7-O9 stars and B0-B3 stars are consistent with both  
distances of 4.8 kpc and 2.4 kpc, the shorter distance would lead to extremely low 
luminosities for the three transitional objects, as shown in Table \ref{tablecl1813}.
For example, for the O6-O7If star,   \Mk=$ -5.87\pm0.11$ mag at 4.8 kpc 
or $ -4.36\pm0.17$ mag at 2.4 kpc; 
the latter value is not compatible with the supergiant class. 
The spectrum of O6-O7If star resembles 
that of star F15 in the Arches cluster \citep{martins08};  
for F15 we derive \Mk=$-5.76$ mag by using the photometry by 
\citet{figer02}, the extinction law by \citet{messineo05}, and a distance of 8.4 kpc. 
Thereby, the O6-O7If star  must be located behind the W33 complex.  

Furthermore, for a distance of 2.4 kpc, the cluster members 
of spectral type O or B   would have initial masses below 25 \Msun, i.e., below the 
theoretical and observed lower mass limit for the progenitors of late WN stars 
\citep{georgy12,messineo11}. As a consequence  it would be difficult to understand the presence 
of the two WN7  \citep[\#4, \#7 in ][]{messineo11}, one O8-9If/WN9h,  and  one O6-O7If star, 
given the resultant  absence of a progenitor population. 
The average \Mk = $-5.86$ mag  ($\sigma=0.43$ mag) inferred for the two WN7, 
cluster members, at 4.8 kpc is consistent with that of WN7 stars ($-5.38$ mag with a $\sigma=0.38$ mag) 
in Westerlund 1 \citep{crowther06wd1}.

\begin{table*}
\caption{\label{tablecl1813} Average \Mk,  \Aks, and number of stars   per spectral group and  luminosity class in Cl 1813-178
\citep[Table2,][]{messineo11}.}
\begin{tabular}{ll|lll|lll|llr}
\hline
\hline
         &        & \multicolumn{3}{c}{\rm Cluster distance$^{(a)}$ 4.8 kpc} &\multicolumn{3}{c}{\rm W33 distance$^{(b)}$ 2.4 kpc}& \\ 
Sp. group& class$^{(c)}$  &                                   \Mk$^{(d)}$     & \Aks$^{(d)}$ & Nstar & \Mk$^{(d)}$& \Aks$^{(d)}$ & Nstar                             \\
\hline
B0-B3    & I      & $-6.32 \pm 1.01$ & $0.94 \pm 0.25$ &  7 &    $-5.78 \pm 0.33$ &$0.87 \pm 0.18$ &  3  &  \\    
B0-B3    & V/III  & $-4.24 \pm 0.60$ & $0.98 \pm 0.34$ &  5 &    $-3.32 \pm 0.90$ &$0.97 \pm 0.30$ &  9  &  \\ 
O7-O9    & I      & $-5.48 \pm 0.28$ & $0.84 \pm 0.09$ &  5 &    $..$             & $..$           &  0  &  \\
O7-O9    & V/III  & $-4.93 \pm 0.11$ & $0.86 \pm 0.02$ &  1 &    $-3.88 \pm 0.36$ &$0.83 \pm 0.09$ &  6  &  \\  
WN7      & I      & $-5.86 \pm 0.43$ & $0.75 \pm 0.08$ &  2 &    $-4.35 \pm 0.43$ &$0.75 \pm 0.08$ &  2  &  \\
O6O7If   & I      & $-5.87 \pm 0.11$ & $1.03 \pm 0.02$ &  1 &    $-4.36 \pm 0.17$ &$1.03 \pm 0.02$ &  1  &  \\
O8O9If   & I      & $-7.32 \pm 0.10$ & $1.24 \pm 0.01$ &  1 &    $-5.81 \pm 0.16$ &$1.24 \pm 0.01$ &  1  &  \\
cLBV     & I      & $-7.53 \pm 0.10$ & $1.09 \pm 0.02$ &  1 &    $-6.03 \pm 0.16$ &$1.09 \pm 0.02$ &  1  &  \\
\hline
\end{tabular}
\begin{list}{}{}{}
\item[{\bf Notes.}]  $^{(a)}$ Cluster kinematic distance \citep{messineo08,reid09}.~
$^{(b)}$ W33 distance \citep{immer13}.~ 
$^{(c)}$ Luminosity classes are photometrically assigned:
for B0-B3 supergiants \Mk\ $ <-5.0 $ mag, for B0-B3 giants or dwarfs  \Mk\ $ >-5.0 $ mag.
For O7-O9 supergiants \Mk\ $ <-5.0 $ mag, for O7-O9 giants or dwarfs  \Mk\ $ >-5.0 $ mag.
 $^{(d)}$ When Nstar $>1$, quoted errors are the standard deviations.
\end{list}
\end{table*}

While the SN that gave rise to the remnant  G12.72$-$0.00 may have occurred in 
Cl 1813$-$178 (given the precise superposition), a physical  association with SNR G12.82$-$0.02 and 
associated pulsar PSR J1813$-$1749 appears doubtful. Specifically, a  comparison of the  significant 
column  density to the pulsar and SNR  to the less extreme extinction  inferred for cluster members  
led \citet{halpern12} to conclude that SNR G12.82$-$0.02 likely lies beyond both the  cluster and the 
W33 complex ($d\sim 5 - 12$ kpc). 

Finally, it is of interest that the distance estimate for  the W33 complex was determined from parallax 
measurement of masers. If it were to be observed in the future, when such emission had ceased, 
it would be difficult to recognize it as a complex of discrete sources at the same distance,
and to further distinguish the distinct stellar population of Cl1813$-$178 that  is
projected on the edge of W33 (see Fig.\ \ref{fullmap}).

\section{Summary}
\label{summary}

We performed a near-IR  spectroscopic 
survey for massive stars, encompassing both W33  and the nearby mid-IR bubble/stellar cluster  N10/BDS2003-115 
to study their star formation history.

\begin{itemize}
\item We detected a total of fourteen  new  early-type (OB and WR) stars and a further nine stars with 
spectra consistent with spectral types earlier than F. 
A large population of giants with spectral types G-M were uncovered, but no cool supergiants associated 
with W33 were identified. 

\item
Following \citet{clark09}, the lack of RSGs precludes substantive 
star formation activity with W33 $\geq6$ Myr, while the detected stellar population appears broadly 
consistent with an age of $\sim 2-4$ Myr. 

\item
The complex contains protostars (most notably the embedded protocluster W33 Main and the 
high mass protostar W33A), massive evolved stars,
and clear marks of sequential star formation and feedback.
Star formation within W33 has not led to the formation of 
rich dense clusters, and the  size of W33 (radius $\approx 5$ pc)
is typical for  loose associations \citep{pfalnzer09}.
When the GMC is exhausted and star formation has  ceased, 
W33  will most likely resemble a loose, non-coeval stellar 
association similar to (but less massive than)  Cyg OB2 
\citep[e.g.,][]{negueruela08}.  

\item
Given the spare nature of individual stellar `aggregates’  and the limitations of the current data,
we cannot infer integrated masses for the young 
populations within  W33. W33 is probable less massive 
than other massive star forming complexes  of the Milky Way with known evolved stars,
such as W43 \citep[e.g.,][]{blum99,chenx}, W51 \citep{clark09}, Carina \citep[e.g.,][]{preibisch11},  
and G305 \citep{davies12}, as suggested by their respective integrated radio and IR luminosities 
\citep[e.g.,][]{immer13,conti04}. 

\item
The greater distance to the nearby young massive stellar aggregate 
Cl 1813$-$178   precludes a  physical association with W33. 
The late O-type and B-type members of the cluster support 
a distinct older population than that observed  in W33.
Considering the extinction of Cl 1813$-$178 (\Av=9.1 mag), 
optical spectroscopy would yield precise spectral-types and direct luminosity 
determinations for its constituent stars \citep[e.g,][]{negueruela10}.

\item
Given the distances to W33 and to Cl 1813$-$178, an 
association with the energetic pulsar PSR J1813$-$1749 appears doubtful.

\end{itemize}

\begin{table*}[u] \renewcommand{\arraystretch}{0.8} 
\begin{center}
\caption{ \label{table.phot} Infrared measurements  of the spectroscopically
detected  stars. Identification numbers are taken from Table \ref{table.obspectra},
\ref{table:obslate}.} 
\begin{tabular}{@{\extracolsep{-.07in}} l|rrr|rrr|rrr|r|rrrr|r|rr|rrrrr}
\hline
    &  \multicolumn{3}{c}{\rm 2MASS}   &\multicolumn{3}{c}{\rm DENIS} &\multicolumn{3}{c}{\rm UKIDSS} &   \multicolumn{4}{c}{\rm GLIMPSE}   &  & \multicolumn{1}{c}{\rm MSX}& \multicolumn{2}{c}{\rm WISE}  & \multicolumn{1}{c}{\rm NOMAD} &\\ 
\hline 
 {\rm ID} & {\it J} & {\it H} & { $K_S$}  &
 {\it I} & {\it J} & { $K_S$} & 
 {\it J} & {\it H} & { $K$} & {\rm Flag} &
 {\rm [3.6]} & {\rm [4.5]} & {\rm [5.8]} & {\rm [8.0]} &
 {\it A}  &{\it W1} &{\it W2}  & {\it R} & \\ 
\hline 
 &{\rm [mag]}   &	{\rm [mag]}    & {\rm [mag]}     & {\rm [mag]} &{\rm [mag]}  &  {\rm [mag]} &{\rm [mag]} 
 &{\rm [mag]}  & {\rm [mag]}&&{\rm [mag]}&{\rm [mag]} &{\rm [mag]}&{\rm [mag]}&{\rm [mag]}& {\rm [mag]}& {\rm [mag]}& {\rm [mag]}& \\ 
\hline 
                             1  &  10.168 &  8.887 &  8.024 & 13.969 & 10.053 &  8.059 & \nodata & \nodata & \nodata &  2  & 7.247 &  6.686 &  6.520 &  6.192 & \nodata & \nodata & \nodata & \nodata &  \\
                             2  &  15.364 & \nodata & \nodata & \nodata & 15.171 & 12.121 & 15.252 & 13.922 & 13.266 &  1  &\nodata & \nodata & \nodata & \nodata & \nodata & \nodata & \nodata & \nodata &  \\
                             3  &  13.230 & 11.658 & 10.913 & \nodata & 13.203 & 10.831 & 13.267 & 11.592 & 10.907 &  1  &10.364 & 10.462 & 10.058 & \nodata & \nodata & 10.631 & 10.170 & \nodata &  \\
                             4  &  11.756 & 10.324 &  9.653 & 16.440 & 11.561 &  9.668 & 11.561 & 10.231 &  9.500 &  2  & 9.008 &  8.893 &  8.686 &  8.493 & \nodata &  9.050 &  8.896 & \nodata &  \\
                             5  &  13.018 & \nodata & \nodata & 14.124 & 13.120 & \nodata & 13.032 & 12.640 & 12.383 &  1  &\nodata & \nodata & \nodata & \nodata & \nodata & \nodata & \nodata &  14.45 &  \\
                             6  &  \nodata & \nodata & \nodata & \nodata & \nodata & \nodata & 11.718 & 10.301 &  9.466 &  1  &\nodata & \nodata & \nodata & \nodata & \nodata & \nodata & \nodata & \nodata&  \\
                             7  &   9.376 &  8.508 &  8.090 & 12.117 &  9.457 &  8.008 &  9.389 & \nodata & \nodata &  2  & 7.737 &  7.764 &  7.661 &  7.730 & \nodata & \nodata & \nodata &  14.07 &  \\
                             8  &  14.913 & 11.581 &  9.892 & \nodata & 14.629 &  9.734 & 15.104 & 11.675 &  9.805 &  2  & 8.588 &  8.296 &  7.955 &  8.126 & \nodata &  8.627 &  8.056 & \nodata &  \\
                             9  &  \nodata & 13.858 & \nodata & \nodata & \nodata & \nodata & 17.322 & 14.083 & 12.377 &  1  &\nodata & \nodata & \nodata & \nodata & \nodata & \nodata & \nodata & \nodata &  \\
                            10  &  \nodata & \nodata & \nodata & \nodata & \nodata & \nodata & 14.697 & 12.416 & 11.185 &  1  &\nodata & \nodata & \nodata & \nodata & \nodata & \nodata & \nodata & \nodata&  \\
                            11  &  \nodata & 11.713 & 10.363 & \nodata & \nodata & 10.249 & 14.325 & 11.703 & 10.278 &  2  & 9.401 &  9.127 &  8.835 & \nodata & \nodata & \nodata & \nodata & \nodata &  \\
                            12  &  \nodata & \nodata & \nodata & 13.372 & 12.081 & \nodata & 12.907 & 12.543 & 12.505 &  1  &\nodata & \nodata & \nodata & \nodata & \nodata & \nodata & \nodata & \nodata&  \\
                            13  &  12.718 & 11.524 & 10.826 & 16.555 & 12.634 & 10.698 & 12.681 & 11.507 & 10.894 &  1  &\nodata & \nodata & \nodata & \nodata & \nodata & \nodata & \nodata & \nodata &  \\
                            14  &  \nodata & \nodata & 12.224 & \nodata & \nodata & \nodata & 16.918 & 14.036 & 12.156 &  1  &10.408 &  9.677 & 10.118 & \nodata & \nodata &  9.428 &  8.128 & \nodata &  \\
                            15  &  13.995 & 13.189 & \nodata & 15.927 & 13.880 & 12.056 & 14.006 & 13.271 & 13.023 &  1  &\nodata & \nodata & \nodata & \nodata & \nodata & \nodata & \nodata &  17.07 &  \\
                            16  &  13.535 & 12.944 & \nodata & 15.439 & 13.374 & \nodata & 13.538 & 12.896 & 12.633 &  1  &\nodata & \nodata & \nodata & \nodata & \nodata & \nodata & \nodata &  16.85 &  \\
                            17  &  15.411 & 13.290 & 12.102 & \nodata & 14.942 & 11.975 & 15.454 & 13.367 & 12.175 &  1  &11.305 & \nodata & \nodata & \nodata & \nodata & \nodata & \nodata & \nodata &  \\
                            18  &  13.457 & 11.652 & 10.567 & 18.408 & 13.375 & 10.523 & 13.526 & 11.648 & 10.603 &  1  &\nodata & \nodata & \nodata & \nodata & \nodata & \nodata & \nodata & \nodata &  \\
                            19  &  14.654 & 13.826 & 13.053 & 16.450 & 14.639 & \nodata & 14.699 & 13.984 & 13.797 &  1  &\nodata & \nodata & \nodata & \nodata & \nodata & \nodata & \nodata &  16.72 &  \\
                            20  &  15.044 & 14.084 & 12.918 & 16.900 & 14.915 & \nodata & 15.502 & 14.881 & 14.638 &  1  &\nodata & \nodata & \nodata & \nodata & \nodata & \nodata & \nodata & \nodata &  \\
                            21  &  13.531 & 12.692 & 11.914 & 15.770 & 13.464 & 11.944 & 13.538 & 12.829 & 12.488 &  1  &\nodata & \nodata & \nodata & \nodata & \nodata & \nodata & \nodata &  17.07 &  \\
                            22  &  15.983 & 12.664 & 10.448 & \nodata & \nodata & 10.569 & 15.421 & 12.369 & 10.072 &  2  & 8.006 &  7.283 &  6.627 &  5.962 & \nodata &  7.868 &  6.766 & \nodata &  \\
                            23  &  10.062 &  8.741 &  8.103 & 14.030 &  9.997 &  8.091 &  9.846 & \nodata &  7.946 &  2  & 7.604 &  7.493 &  7.466 & \nodata & \nodata & \nodata & \nodata &  16.65 &  \\
                            24  &  12.724 & 11.505 & 10.986 & 15.478 & 12.838 & 10.989 & 12.724 & 11.500 & 11.036 &  1  &\nodata & \nodata & \nodata & \nodata & \nodata & \nodata & \nodata &  17.10 &  \\
                            25  &  \nodata & 13.023 & 10.989 & \nodata & \nodata & 10.989 & 17.218 & 12.996 & 10.970 &  1  & 9.636 &  9.516 &  9.148 &  9.356 & \nodata &  9.342 &  9.077 & \nodata &  \\
                            26  &  10.318 &  9.128 &  8.607 & 13.144 & 10.164 &  8.545 & 10.278 & \nodata &  8.652 &  2  & 8.323 &  8.362 &  8.160 &  8.200 & \nodata &  8.425 &  8.419 &  15.18 &  \\
                            27  &  12.089 & 10.985 & 10.546 & 14.682 & 11.923 & 10.514 & 12.059 & 10.960 & 10.592 &  2  &10.232 & 10.385 &  9.741 &  9.343 & \nodata & 10.251 &  9.900 &  16.17 &  \\
                            28  &  \nodata & 11.343 & 10.069 & \nodata & 14.130 & 10.280 & \nodata & \nodata & \nodata &  2  & 9.197 &  9.273 &  8.896 &  8.929 & \nodata &  9.232 &  9.126 & \nodata&  \\
                            29  &  \nodata & 13.950 & 12.888 & \nodata & \nodata & \nodata & 16.656 & 14.289 & 13.069 &  3  &12.040 & 12.072 & 11.984 & \nodata & \nodata & \nodata & \nodata & \nodata&  \\
                            30  &  14.448 & 11.369 &  9.839 & \nodata & 14.534 &  9.825 & \nodata & \nodata & \nodata &  2  & 8.694 &  8.764 &  8.320 &  8.301 & \nodata &  9.002 &  8.648 & \nodata&  \\
                            31  &  \nodata & 12.221 &  9.892 & \nodata & \nodata &  9.827 & 16.645 & 12.221 &  9.796 &  2  & 8.197 &  8.104 &  7.627 &  7.768 & \nodata &  8.452 &  8.040 & \nodata &  \\
                            32  &  14.969 & 11.788 & \nodata & \nodata & 14.681 & 10.123 & 15.392 & 11.835 & 10.081 &  2  &\nodata & \nodata & \nodata & \nodata & \nodata & \nodata & \nodata & \nodata &  \\
                            33  &  \nodata & 13.074 & 10.931 & \nodata & \nodata & 11.053 & 17.542 & 13.155 & 10.921 &  1  & 9.415 &  9.390 &  8.896 &  9.114 & \nodata &  9.651 &  9.325 & \nodata &  \\
                            34  &  13.342 & \nodata & \nodata & 15.133 & 13.105 & 10.772 & 13.464 & 12.651 & 12.363 &  1  &\nodata & \nodata & \nodata & \nodata & \nodata & \nodata & \nodata &  16.69 &  \\
                            35  &  \nodata & \nodata & \nodata & \nodata & \nodata & \nodata & \nodata & 14.043 & 11.403 &  1  &\nodata & \nodata & \nodata & \nodata & \nodata &  9.429 &  9.049 & \nodata&  \\
                            36  &  \nodata & \nodata & \nodata & \nodata & \nodata & \nodata & \nodata & 15.959 & 13.018 &  1  &\nodata & \nodata & \nodata & \nodata & \nodata & \nodata & \nodata & \nodata&  \\
                            37  &  \nodata & 12.807 & 10.785 & \nodata & \nodata & \nodata & 16.846 & 13.052 & 10.779 &  1  & 9.256 &  9.128 &  8.682 &  8.898 & \nodata &  9.315 &  8.845 &  17.20 &  \\
                            38  &  \nodata & 12.477 & 10.530 & \nodata & \nodata & 10.431 & \nodata & 12.532 & 10.516 &  1  & 9.111 &  9.032 &  8.567 &  8.698 & \nodata &  9.363 &  9.007 & \nodata &  \\
                            39  &  12.522 & 11.127 & 10.530 & 16.098 & 12.530 & 10.516 & 12.461 & 11.216 & 10.489 &  3  &10.056 & 10.088 &  9.952 & \nodata & \nodata & 10.334 & 10.153 & \nodata &  \\
                            40  &  12.236 & 11.290 & 10.841 & \nodata & \nodata & 10.802 & 12.283 & 11.310 & 10.965 &  2  &\nodata & \nodata & \nodata & \nodata & \nodata & \nodata & \nodata & \nodata &  \\
                            41  &  \nodata & 12.569 & 10.368 & \nodata & \nodata & 10.386 & 17.008 & 12.560 & 10.313 &  2  & 8.750 &  8.632 &  8.171 &  8.274 & \nodata &  8.903 &  8.198 & \nodata &  \\
                            42  &  13.305 & 10.899 &  9.761 & \nodata & \nodata &  9.855 & \nodata & \nodata & \nodata &  2  & 8.978 &  9.152 &  8.785 &  8.892 & \nodata &  9.042 &  9.091 & \nodata&  \\
                            43  &  12.808 & 10.941 & 10.124 & 17.603 & 12.943 & 10.083 & \nodata & \nodata & \nodata &  2  & 9.557 &  9.538 &  9.405 &  9.470 & \nodata &  9.595 &  9.665 & \nodata&  \\
                            44  &  13.968 & 11.433 & 10.026 & \nodata & 14.263 & 10.288 & \nodata & \nodata & \nodata &  2  & 8.799 &  8.553 &  8.178 &  7.981 & \nodata &  8.606 &  8.202 & \nodata&  \\
                            45  &  14.392 & 12.009 & 10.893 & \nodata & 14.597 & 11.000 & 14.308 & 12.033 & 10.879 &  3  &10.193 & 10.215 &  9.907 &  9.945 & \nodata & \nodata & \nodata & \nodata&  \\
                            46  &  12.919 & 11.298 & 10.593 & 17.500 & 12.991 & 10.714 & 12.757 & 11.432 & 10.607 &  3  &10.111 & 10.191 &  9.971 &  9.888 & \nodata & 10.195 & 10.227 & \nodata&  \\
                            47  &  11.302 &  8.275 &  6.604 & \nodata & 12.227 &  7.338 & \nodata & \nodata & \nodata &  2  & 5.859 &  6.023 &  4.934 &  4.659 &  4.173 &  5.914 &  4.960 & \nodata&  \\
                            48  &  \nodata & 12.102 & \nodata & \nodata & \nodata & 10.448 & 15.308 & 12.232 & 10.695 &  3  &\nodata & \nodata & \nodata & \nodata & \nodata & \nodata & \nodata & \nodata&  \\

\hline
\end{tabular}
\begin{list}{}{}
\item[{\bf Notes.}] The Flag column indicates which  $JHK$ was adopted in the paper:
Flag=0 source missing; Flag=1 UKIDSS psf-fitting magnitudes; Flag=2   2MASS magnitudes;
Flag=3 UKIDSS DR6 release.
\end{list}
\end{center}
\end{table*}

\addtocounter{table}{-1} 
\begin{table*}[u] \renewcommand{\arraystretch}{0.8} 
\begin{center}
\caption{ Continuation of Table \ref{table.phot}. } 
\begin{tabular}{@{\extracolsep{-.07in}} l|rrr|rrr|rrr|r|rrrr|r|rr|rrrrr}
\hline 
    &  \multicolumn{3}{c}{\rm 2MASS}   &\multicolumn{3}{c}{\rm DENIS} &\multicolumn{3}{c}{\rm UKIDSS} &   \multicolumn{4}{c}{\rm GLIMPSE}   & & \multicolumn{1}{c}{\rm MSX}& \multicolumn{2}{c}{\rm WISE}  & \multicolumn{1}{c}{\rm NOMAD} &\\ 
\hline 
 {\rm ID} & {\it J} & {\it H} & { $K_S$}  &
 {\it I} & {\it J} & { $K_S$} & 
 {\it J} & {\it H} & { $K$} & {\rm Flag} &
 {\rm [3.6]} & {\rm [4.5]} & {\rm [5.8]} & {\rm [8.0]} &
 {\it A}  &{\it W1} &{\it W2}  & {\it R} & \\ 
\hline 
 &{\rm [mag]}   &	{\rm [mag]}    & {\rm [mag]}     & {\rm [mag]} &{\rm [mag]}  &  {\rm [mag]} &{\rm [mag]} 
 &{\rm [mag]}  & {\rm [mag]}&&{\rm [mag]}&{\rm [mag]} &{\rm [mag]}&{\rm [mag]}&{\rm [mag]}& {\rm [mag]}& {\rm [mag]}& {\rm [mag]}& \\ 
\hline 
                            49  &  \nodata & 12.343 & 10.054 & \nodata & \nodata &  9.894 & 16.907 & 12.242 &  9.911 &  2  & 8.469 &  8.309 &  7.934 &  8.103 & \nodata &  8.623 &  8.238 & \nodata &  \\
                            50  &  \nodata & 12.402 &  9.991 & \nodata & \nodata &  9.864 & 17.441 & 12.554 &  9.971 &  2  & 8.185 &  7.804 &  7.372 &  7.380 & \nodata &  8.345 &  7.507 & \nodata &  \\
                            51  &  15.406 & \nodata & \nodata & \nodata & 15.207 &  8.948 & 15.483 & 11.260 &  9.029 &  1  &\nodata & \nodata & \nodata & \nodata & \nodata & \nodata & \nodata & \nodata &  \\
                            52  &  \nodata & 13.189 & 10.908 & \nodata & \nodata & 11.030 & 18.233 & 13.486 & 11.088 &  1  & 9.378 &  9.157 &  8.802 &  8.914 & \nodata &  9.656 &  9.153 & \nodata &  \\
                            53  &  \nodata & 12.879 & 10.554 & \nodata & \nodata & 10.341 & 17.624 & 12.909 & 10.466 &  1  & 8.842 &  8.678 &  8.213 &  8.258 & \nodata &  9.192 &  8.700 & \nodata &  \\
                            54  &  \nodata & 11.876 &  9.650 & \nodata & 16.250 &  9.491 & 16.570 & 11.990 &  9.599 &  2  & 8.104 &  7.997 &  7.471 &  7.575 & \nodata &  8.038 &  7.608 & \nodata &  \\
                            55  &  \nodata & 14.413 & 12.678 & \nodata & \nodata & \nodata & 18.239 & 14.487 & 12.653 &  1  &11.443 & 11.259 & 11.026 & \nodata & \nodata & \nodata & \nodata & \nodata &  \\
                            56  &  10.359 &  8.796 &  8.074 & 14.426 & 10.261 &  8.171 & 10.272 & \nodata &  8.390 &  2  & 7.647 &  7.861 &  7.551 &  7.584 & \nodata &  7.708 &  7.902 &  18.21 &  \\
                            57  &  \nodata & 10.797 &  8.532 & \nodata & \nodata & \nodata & 14.561 & 10.353 & \nodata &  2  & 6.713 &  5.976 &  5.456 &  5.432 &  5.965 &  6.669 &  5.790 & \nodata&  \\
                            58  &  \nodata & \nodata & \nodata & \nodata & \nodata & \nodata & \nodata & 15.267 & 13.008 &  1  &\nodata & \nodata & \nodata & \nodata & \nodata & \nodata & \nodata & \nodata&  \\
                            59  &  \nodata & \nodata & \nodata & \nodata & \nodata & \nodata & \nodata & \nodata & 11.213 &  1  & 8.730 &  8.409 &  7.943 &  7.922 & \nodata & \nodata & \nodata & \nodata&  \\
                            60  &  10.983 & 10.173 &  9.733 & 12.567 & 11.113 &  9.816 & 10.938 & 10.119 &  9.976 &  2  &\nodata & \nodata & \nodata & \nodata & \nodata & \nodata & \nodata &  12.38 &  \\
                            61  &  \nodata & 11.646 &  9.500 & \nodata & \nodata &  9.332 & 16.123 & 11.743 &  9.392 &  2  & 7.871 &  7.812 &  7.317 &  7.462 & \nodata &  8.263 &  7.906 & \nodata&  \\
                            62  &  \nodata & \nodata & \nodata & \nodata & \nodata & \nodata & 15.359 & 11.039 &  8.877 &  1  &\nodata & \nodata & \nodata & \nodata & \nodata &  7.684 &  7.327 & \nodata&  \\
                            63  &  \nodata & 13.134 & 10.946 & \nodata & \nodata & 10.815 & 17.622 & 13.143 & 10.890 &  1  & 9.412 &  9.277 &  8.807 &  9.079 & \nodata &  9.682 &  9.328 & \nodata &  \\
                            64  &  \nodata & \nodata & \nodata & \nodata & \nodata & \nodata & \nodata & 15.078 & 13.038 &  1  &\nodata & \nodata & \nodata & \nodata & \nodata & \nodata & \nodata & \nodata&  \\
                            65  &  \nodata & 12.565 & 10.516 & \nodata & \nodata & \nodata & 16.647 & 12.667 & 10.534 &  1  & 9.065 &  8.912 &  8.525 &  8.784 & \nodata &  9.266 &  8.899 & \nodata &  \\
                            66  &  \nodata & \nodata & \nodata & \nodata & \nodata & \nodata & \nodata & 15.982 & 13.826 &  1  &\nodata & \nodata & \nodata & \nodata & \nodata & \nodata & \nodata & \nodata&  \\
                            67  &  15.249 & 11.288 &  9.247 & \nodata & 15.581 &  9.271 & 15.312 & 11.562 &  9.341 &  2  & 7.967 &  7.485 &  7.035 &  7.010 & \nodata &  8.355 &  7.534 & \nodata &  \\
                            68  &  \nodata & \nodata & 10.766 & \nodata & \nodata & 10.654 & 16.547 & 12.617 & 10.705 &  1  & 9.434 &  9.310 &  8.951 &  8.993 & \nodata & \nodata & \nodata & \nodata &  \\
                            69  &  \nodata & \nodata & 12.400 & \nodata & \nodata & \nodata & 15.949 & 13.533 & 12.488 &  1  &11.679 & 11.694 & \nodata & \nodata & \nodata & \nodata & \nodata & \nodata &  \\
                            70  &   9.863 &  8.703 &  8.271 & 12.260 &  9.801 &  8.243 &  9.689 & \nodata &  8.248 &  2  &\nodata & \nodata & \nodata & \nodata & \nodata & \nodata & \nodata &  14.59 &  \\
                            71  &  12.300 &  9.616 &  8.313 & \nodata & 12.316 &  8.424 & \nodata & \nodata & \nodata &  2  & 7.347 &  7.563 &  7.112 &  7.140 & \nodata &  7.551 &  7.476 & \nodata&  \\
                            72  &  \nodata & \nodata & \nodata & \nodata & \nodata & \nodata & 17.829 & 13.790 & 11.754 &  1  &\nodata & \nodata & \nodata & \nodata & \nodata & 10.010 &  9.864 & \nodata&  \\
                            73  &  13.698 & 11.808 & 10.830 & \nodata & \nodata & \nodata & 13.851 & 11.891 & 11.029 &  1  &\nodata & \nodata & \nodata & \nodata & \nodata & \nodata & \nodata & \nodata&  \\
                            74  &  \nodata & 13.951 & \nodata & \nodata & \nodata & \nodata & 16.528 & 14.797 & 13.974 &  1  &\nodata & \nodata & \nodata & \nodata & \nodata & \nodata & \nodata & \nodata &  \\
                            75  &  \nodata & \nodata & \nodata & \nodata & \nodata & \nodata & \nodata & 15.863 & 14.412 &  1  &\nodata & \nodata & \nodata & \nodata & \nodata & \nodata & \nodata & \nodata&  \\
                            76  &   9.682 &  8.697 &  8.351 & 11.487 &  9.656 &  8.361 & \nodata & 10.723 &  8.752 &  2  & 8.216 &  8.268 &  8.202 & \nodata & \nodata &  7.849 &  7.762 &  13.15 &  \\
                            77  &  15.058 & 12.899 & 11.936 & \nodata & 15.116 & 11.803 & 15.083 & 12.965 & 12.026 &  1  &11.369 & 11.088 & 10.888 & 11.088 & \nodata & \nodata & \nodata & \nodata &  \\
                            78  &  14.697 & 12.579 & 11.684 & \nodata & 14.524 & 11.579 & 14.617 & 12.601 & 11.698 &  1  &11.050 & 10.898 & 10.814 & 10.222 & \nodata & \nodata & \nodata & \nodata &  \\
                            79  &  14.802 & 12.587 & 11.665 & \nodata & 14.527 & 11.717 & 14.759 & 12.634 & 11.679 &  1  &11.028 & 10.899 & 10.928 & \nodata & \nodata & \nodata & \nodata & \nodata &  \\
                            80  &  13.517 & 12.010 & 10.395 & 15.174 & 13.668 & 10.284 & 13.497 & 12.047 & 10.353 &  1  & 9.094 &  9.046 &  8.599 &  8.595 & \nodata & \nodata & \nodata & \nodata &  \\
                            81  &  15.107 & 12.377 & 11.174 & \nodata & 14.763 & 11.066 & 15.158 & 12.430 & 11.195 &  1  &10.278 & 10.134 &  9.907 & \nodata & \nodata & \nodata & \nodata & \nodata &  \\
                            82  &  \nodata & \nodata & \nodata & \nodata & \nodata & \nodata & \nodata & \nodata & \nodata &  0  &\nodata & \nodata & \nodata & \nodata & \nodata & \nodata & \nodata & \nodata&  \\
                            83  &  10.041 &  7.668 &  6.538 & 16.385 &  9.916 &  6.406 & \nodata & \nodata & \nodata &  2  & 6.055 &  6.298 &  5.556 &  5.655 & \nodata &  5.821 &  5.700 &  19.11 &  \\
                            84  &   8.914 &  7.061 &  6.169 & 15.116 &  9.053 &  6.262 & \nodata & \nodata & \nodata &  2  & 5.380 & \nodata &  5.099 &  4.903 &  4.824 &  5.529 &  5.257 & \nodata &  \\
                            85  &   9.217 &  7.221 &  6.256 & 15.880 &  9.132 &  6.098 & \nodata & \nodata & \nodata &  2  & 7.037 &  6.028 &  5.355 &  5.246 &  4.897 &  5.642 &  5.308 & \nodata &  \\
                            86  &  10.410 &  7.856 &  6.586 & \nodata & 10.366 &  6.539 & \nodata & \nodata & \nodata &  2  & 5.917 &  6.072 &  5.306 &  5.223 &  5.302 &  5.825 &  5.388 & \nodata &  \\
                            87  &  13.478 &  9.616 &  7.686 & \nodata & 13.446 &  7.654 & \nodata & \nodata & \nodata &  2  & 6.472 &  6.093 &  5.550 &  5.473 &  5.406 &  6.384 &  5.869 & \nodata &  \\
                            88  &  15.278 & 12.589 & 10.172 & 16.571 & 15.264 & 10.091 & \nodata & \nodata & \nodata &  2  & 8.484 &  8.273 &  7.760 &  7.800 & \nodata &  8.891 &  8.200 & \nodata &  \\
                            89  &  \nodata & \nodata & 11.464 & \nodata & \nodata & 11.362 & \nodata & 14.940 & 11.465 &  1  &\nodata & \nodata & \nodata & \nodata & \nodata & \nodata & \nodata & \nodata &  \\
                            90  &  \nodata & 12.981 & 11.020 & \nodata & \nodata & \nodata & 17.045 & 13.036 & 11.008 &  1  & 9.590 &  9.522 &  9.172 & \nodata & \nodata & \nodata & \nodata & \nodata &  \\
                            91  &  14.749 & 12.311 & 11.216 & \nodata & 14.528 & 11.346 & 14.767 & 12.317 & 11.228 &  1  &10.451 & 10.450 & 10.422 & \nodata & \nodata & 10.482 & 10.522 & \nodata &  \\
                            92  &  13.001 & 11.266 & 10.557 & 17.439 & 12.942 & 10.519 & 13.328 & 11.500 & 10.690 &  1  &10.082 & 10.117 & 10.129 & 10.333 & \nodata & 10.022 &  9.973 &  19.88 &  \\
                            93  &  \nodata & \nodata & \nodata & \nodata & \nodata & \nodata & 17.757 & 13.992 & 12.169 &  1  &\nodata & \nodata & \nodata & \nodata & \nodata & \nodata & \nodata & \nodata&  \\

 \hline
\end{tabular}
\begin{list}{}{}
\item[{\bf Notes.}] Stars \#59 and \#60 are blended even at the UKIDSS resolution, 
listed measurements should be taken as upper limits.
\end{list}
\end{center}
\end{table*}

\appendix{
\section{Finding charts }

Figure \ref{ukchart} displays charts for the spectroscopically observed
stars. For a few stars, which are not easily identifiable in this
Fig.\, additional SINFONI charts are provided in Fig.\
\ref{sinfochart}.

\begin{figure*}
\resizebox{0.8\hsize}{!}{\includegraphics[angle=0]{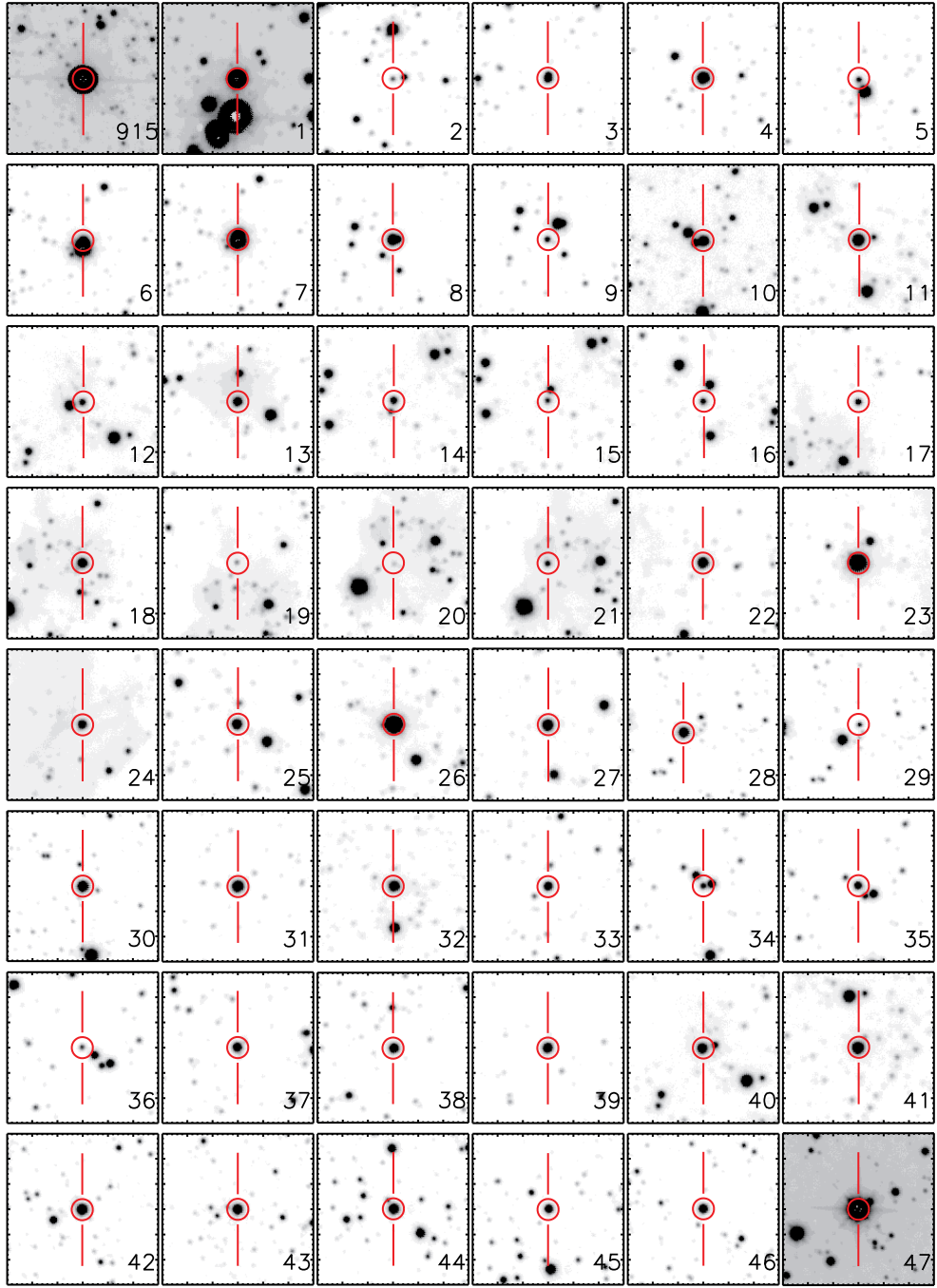}}
\caption{ \label{ukchart}  UKIDSS $K$-band charts ($0\rlap{.}^{\prime}5 \times 0\rlap{.}^{\prime}5$ ) 
of stars spectroscopically  observed. North is up and Est to the left.} 
\end{figure*}

\addtocounter{figure}{-1}
\begin{figure*}
\resizebox{0.8\hsize}{!}{\includegraphics[angle=0]{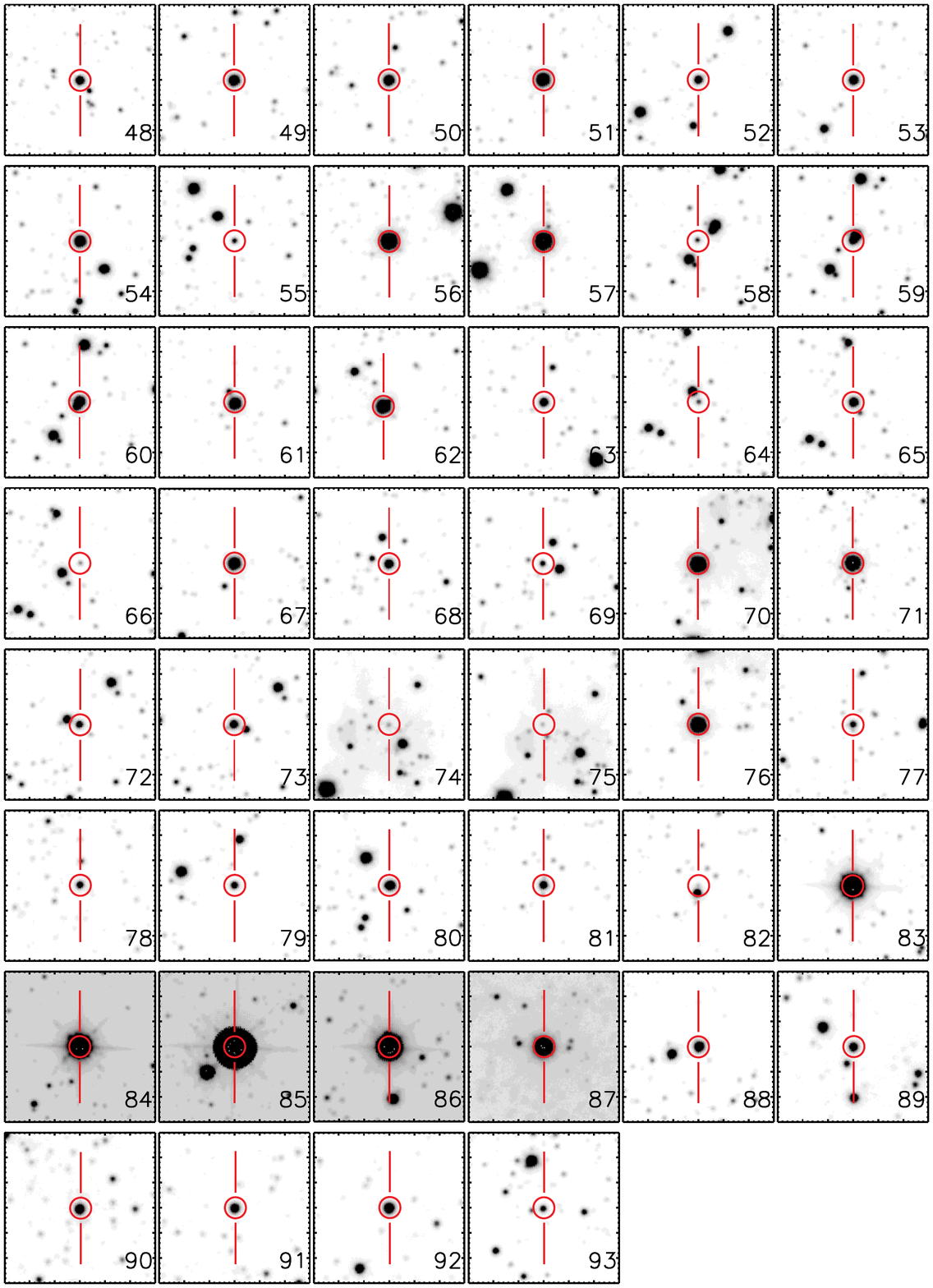}}
\caption{ Continuation of Figure \ref{ukchart}. }
\end{figure*}

\begin{figure*}
\resizebox{0.7\hsize}{!}{\includegraphics[angle=0]{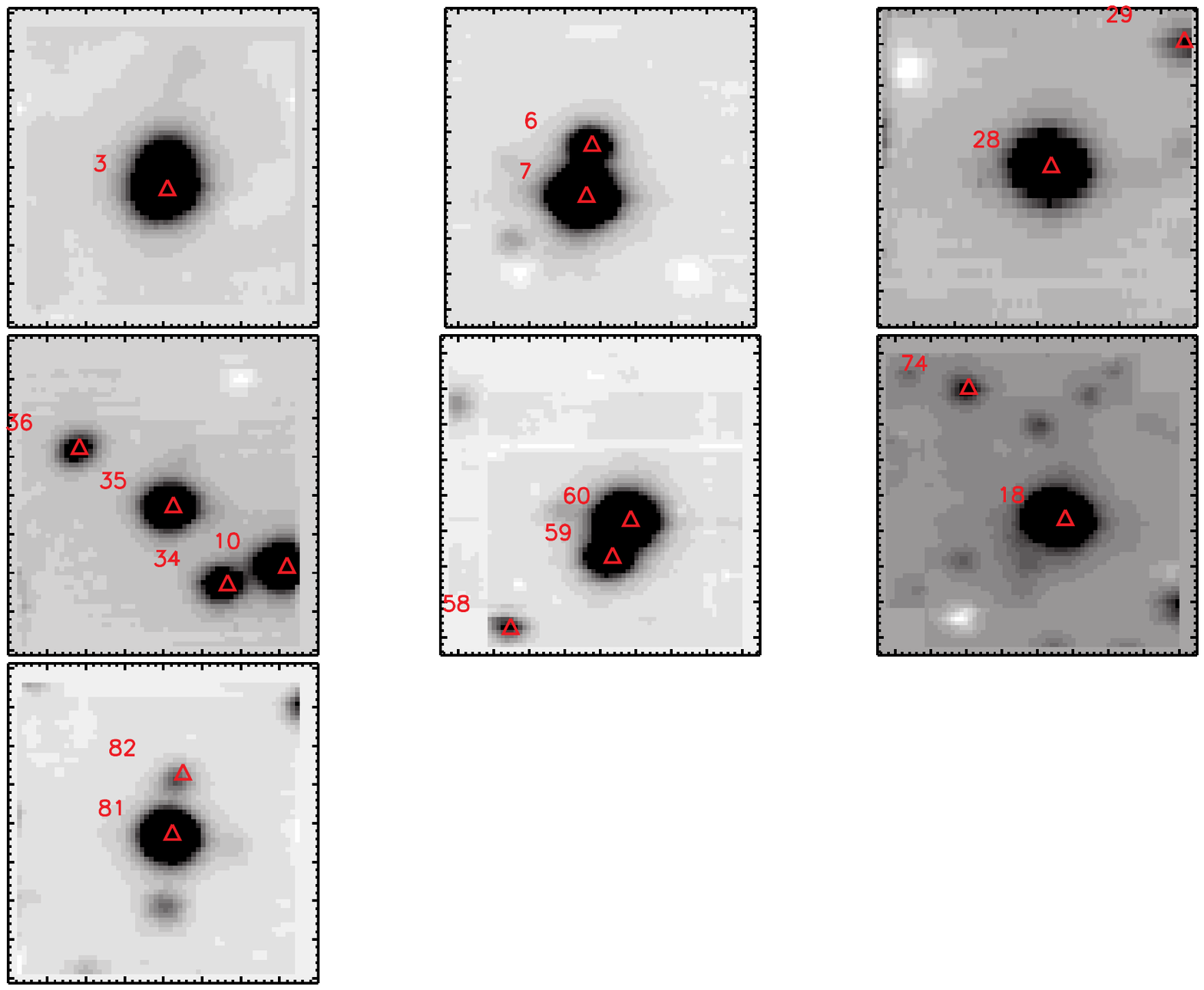}}
\caption{ \label{sinfochart} Additional charts for sources not easily identifiable in Figure \ref{ukchart}.
The displayed images are obtained by averaging the SINFONI 
data-cubes in wavelength ($K$ grating, $8\rlap{}^{\prime\prime} \times 8\rlap{}^{\prime\prime}$). }
\end{figure*}

}

\acknowledgments {
This work was partially funded by the ERC Advanced Investigator Grant GLOSTAR (247078).
FN acknowledges funding from the Spanish Government Ministerio de
Economia y Competitividad (MINECO) through grants AYA2010-21697-C05-01,
FIS2012-39162-C06-01 and ESP2013-47809-C3-1-R.
This publication makes use of data products from the Two Micron All Sky Survey, which is a joint project 
of the University of Massachusetts and the Infrared Processing and Analysis Center/California Institute 
of Technology, funded by the National Aeronautics and Space Administration and the National Science Foundation.
This work is based  on observations made with the Spitzer Space Telescope, 
which is operated by the Jet Propulsion Laboratory, California Institute of Technology under a contract with NASA.
DENIS is  a joint effort of several Institutes mostly located in Europe. It has
    been supported mainly by the French Institut National des
    Sciences de l'Univers, CNRS, and French Education Ministry, the
    European Southern Observatory, the State of Baden-Wuerttemberg, and
    the European Commission under networks of the SCIENCE and Human
    Capital and Mobility programs, the Landessternwarte, Heidelberg and
    Institut d'Astrophysique de Paris.
This research made use of data products from the
Midcourse Space Experiment, the processing of which was funded by the Ballistic Missile Defence Organization with additional
support from the NASA office of Space Science. 
This publication makes use of data products from
WISE, which is a joint project of the University of California, Los
Angeles, and the Jet Propulsion Laboratory/California Insti-
tute of Technology, funded by the National Aeronautics and
Space Administration. 
This work is based in part on data obtained as part of the UKIRT Infrared Deep Sky
Survey. 
This research made use of Montage, funded by the National Aeronautics and Space Administration's 
Earth Science Technology Office, Computational Technnologies Project, under Cooperative 
Agreement Number NCC5-626 between NASA and the California Institute of Technology. 
The code is maintained by the NASA/IPAC Infrared Science Archive.
This research has made use of the VizieR catalogue access tool, CDS, Strasbourg, France
This research has made use of the SIMBAD data base, 
operated at CDS, Strasbourg, France. 
This research has made use of NASA's Astrophysics Data System Bibliographic Services.
A special thank goes to the great support offered by  the European Southern Observatory.
MM thanks the Jos de Bruine and  Timo Prusti for 
useful discussions and support while at ESA.
We thank the referee Dr. Philip Dufton for his careful reading.
}


\end{document}